\documentclass[preprint,12pt]{elsarticle}

\usepackage{amssymb}
\usepackage{amsmath}
\usepackage{amsthm}
\usepackage{xurl}
\usepackage{subcaption}
\usepackage{booktabs}
\usepackage{upgreek}
\usepackage{graphicx}
\usepackage{multirow}
\usepackage{makecell}
\usepackage{arydshln}
\usepackage{xcolor}

\journal{NIM-A}

\begin{document}

\begin{frontmatter}

\title{Further Characterization of the JadePix-3 CMOS Pixel Sensor for the CEPC Vertex Detector: in Dependence of Substrate Reverse Bias}

\author[inst1,inst2]{Jiahao Hu}
\author[inst1,inst2]{Ruiyang Zhang}
\author[inst1,inst2]{Zhiliang Chen}
\author[inst3]{Yunpeng Lu\corref{cor1}}
\ead{yplu@ihep.ac.cn}
\author[inst3,inst4]{Qun Ouyang}
\author[inst1,inst2]{Lailin Xu\corref{cor1}}
\ead{lailinxu@ustc.edu.cn}
\affiliation[inst1]{
  organization={State Key Laboratory of Particle Detection and Electronics, University of Science and Technology of China},
  city={Hefei},
  postcode={230026},
  state={Anhui},
  country={China}
}
\affiliation[inst2]{
  organization={Department of Modern Physics, University of Science and Technology of China},
  city={Hefei},
  postcode={230026},
  state={Anhui},
  country={China}
}
\affiliation[inst3]{
  organization={State Key Laboratory of Particle Detection and Electronics, Institute of High Energy Physics, Chinese Academy of Sciences},
  city={Beijing},
  postcode={100049},
  country={China}
}
\affiliation[inst4]{
  organization={University of Chinese Academy of Sciences},
  city={Beijing},
  postcode={100049},
  country={China}
}
\cortext[cor1]{Corresponding author.}
\begin{abstract}
The Circular Electron-Positron Collider (CEPC), a proposed next-generation $e^+e^-$ collider to enable high-precision studies of the Higgs boson and potential new physics, imposes rigorous demands on detector technologies, particularly the vertex detector. 
JadePix-3 is a prototype Monolithic Active Pixel Sensor (MAPS) designed for the CEPC vertex detector. 
This paper presents a detailed laboratory-based characterization of the JadePix-3 sensor, focusing on the previously under-explored effects of substrate reverse bias voltage on key performance metrics: charge collection efficiency, average cluster size, and hit efficiency of laser. 
Systematic testing demonstrated that JadePix-3 operates reliably under reverse bias, exhibiting a reduced input capacitance, an expanded depletion region, enhanced charge collection efficiency, and a lower fake-hit rate. 
These findings confirm the sensor's potential for high-precision particle tracking and vertexing at the CEPC while offering valuable references for future iterational R\&D of the JadePix series.
\end{abstract}

\begin{highlights}
\item Comprehensive laboratory characterization of the JadePix-3 CMOS pixel prototype for the vertex detector of the Circular Electron-Positron Collider
\item Measurements performed under various reverse bias voltages from 0 to –6~V with dedicated working point optimization at each bias voltage
\item Substrate reverse bias significantly improves the sensor performance such as the charge collection efficiency and the fake-hit rate
\end{highlights}

\begin{keyword}
CEPC \sep Monolithic active pixel sensors \sep CMOS \sep Particle detection
\end{keyword}

\end{frontmatter}



\section{Introduction}
Following the discovery of the Higgs boson, circular $e^+e^-$ colliders have garnered significant interest as premier facilities for high-precision studies of the Standard Model and searches for new physics. The Circular Electron-Positron Collider (CEPC)~\cite{cepccdr, cepctdr} has been proposed to advance this frontier, with an ambitious physics program centered on detailed investigations of the Higgs boson.

The ambitious physics objectives of the CEPC impose stringent performance requirements on its detector systems, particularly on the vertex detector located nearest to the interaction point. As outlined in Table~\ref{tab:vertex}, the vertex detector must satisfy a challenging combination of specifications: exceptional spatial resolution, a minimal material budget, low power consumption, high hit-rate capability, and robust radiation tolerance against beam-induced backgrounds. Meeting these multifaceted demands is a significant technological challenge. In this context, Monolithic Active Pixel Sensors (MAPS) stand out as a uniquely promising technology capable of fulfilling all these criteria~\cite{maps_first_paper, cerncmos}. Consequently, considerable research and development efforts are underway to advance monolithic sensor prototypes for the CEPC. The JadePix sensor series, developed specifically for this purpose, is a direct result of these efforts. This paper focuses on the JadePix-3 prototype, the design specifications of which are briefly summarized below, with further details available in Ref.~\cite{jadepix3}.

\begin{table}[h]
	\caption{Requirements for the CEPC vertex detector~\cite{cepccdr}.}
	\centering
	\begin{tabular}{ll}
		\toprule
            \midrule
		Single point resolution &  3~$\rm{\upmu m}$  \\
		Thickness of sensor chips &  50~$\rm{\upmu m}$  \\
		Power consumption &   \textless 50~$\rm mW/cm^2$  \\
            Timestamp precision &   1~$\rm{\upmu s}$ \\
            Pixel-hit rate &  40~$\rm MHz/cm^2$  \\
            Radiation tolerance &  TID,~3.4 Mrad/year  \\
            ~                   &  NIEL,~$\rm 6.2 \times 10^{12}~\text{1~MeV}~n_{eq}/(cm^2 \cdot year)$\\
		\bottomrule
	\end{tabular}
	\label{tab:vertex}
\end{table}

\subsection{Overview of JadePix-3 design}
\begin{figure}[h]
\centering
\includegraphics[width=0.9\columnwidth]{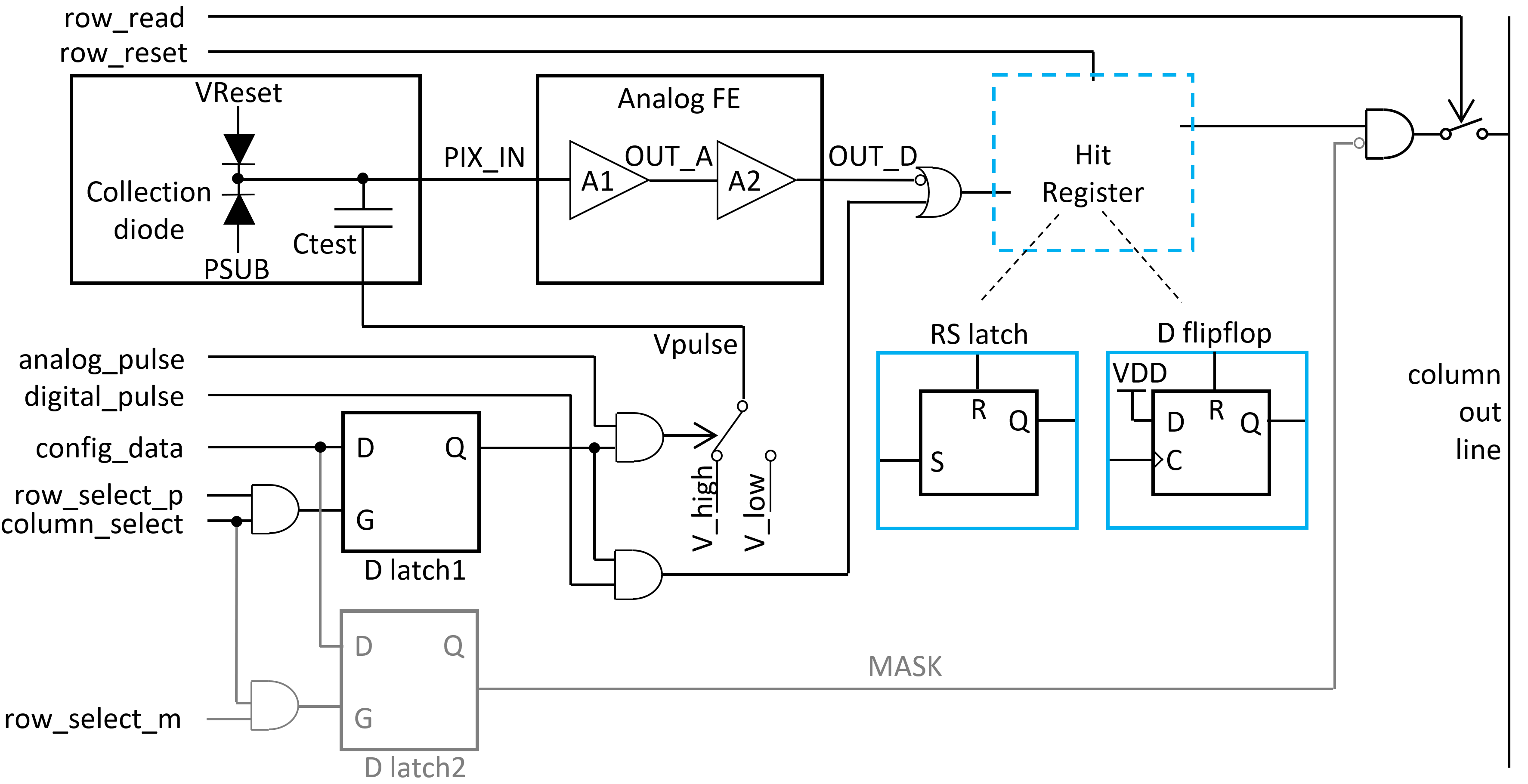}
\caption{Pixel Schematic of the JadePix-3 sensor, illustrating the main components: the collection diode, the analog front-end and the digital readout logic. The digital logic includes configuration bits for electrical pulse test (D latch1) and pixel masking (D latch2). The ``Hit Register'' (blue dashed box) represents the different implementations (RS latch or D flipflop) across the sensor sector designs, as detailed in Tab.~\ref{tab:digital}. Further details on the logic operation can be found in Ref.~\cite{jadepix3} (Section~2.3).}
\label{fig:pixel_schematic}
\end{figure}

The JadePix-3 sensor, developed with TowerJazz 180~nm CMOS imaging technology, covers a pixel matrix consisting of 512 rows and 192 columns. The sensor is divided into four distinguishable sectors in the column direction. While the row pitch is uniformly set at 16 $\rm{\upmu m}$ across all four sectors, the column pitch varies between 23.11 $\rm{\upmu m}$ in sector 2 and 26 $\rm{\upmu m}$ in sectors 0, 1, and 3. The variation in the column pitch arises from the implementation of different digital logic blocks within the pixels. These blocks were designed to evaluate the performance of RS latches versus D flipflops, as well as the inclusion or exclusion of a MASK configuration register, aiming to minimize the pixel layout. The functional block diagram of the pixel is illustrated in Fig.~\ref{fig:pixel_schematic}, while the key differences between the sectors are summarized in Tab.~\ref{tab:digital}.

\begin{table}[h]
	\caption{Digital logic blocks for different sectors.}
	\centering
        \resizebox{\textwidth}{!}{
	\begin{tabular}{llll}
		\toprule
		Sector     & pixel layout     & digital logic block & Remarks \\
		\midrule
		0 & 16 $\times$ 26 $\rm{\upmu m^2}$  & DGT\_V0  & RS latches   \\
		1 & 16 $\times$ 26 $\rm{\upmu m^2}$  & DGT\_V1  & D flipflop, w/o MASK      \\
		2 & 16 $\times$ 23.11 $\rm{\upmu m^2}$  & DGT\_V2  &  RS latches, w/o MASK \\
            3 & 16 $\times$ 26 $\rm{\upmu m^2}$  & DGT\_V0  & Same as Sector 0 but with increased current for the analog front-end \\
		\bottomrule
	\end{tabular}
        }
	\label{tab:digital}
\end{table}

\begin{figure}[h]
\centering
\includegraphics[width=0.7\columnwidth]{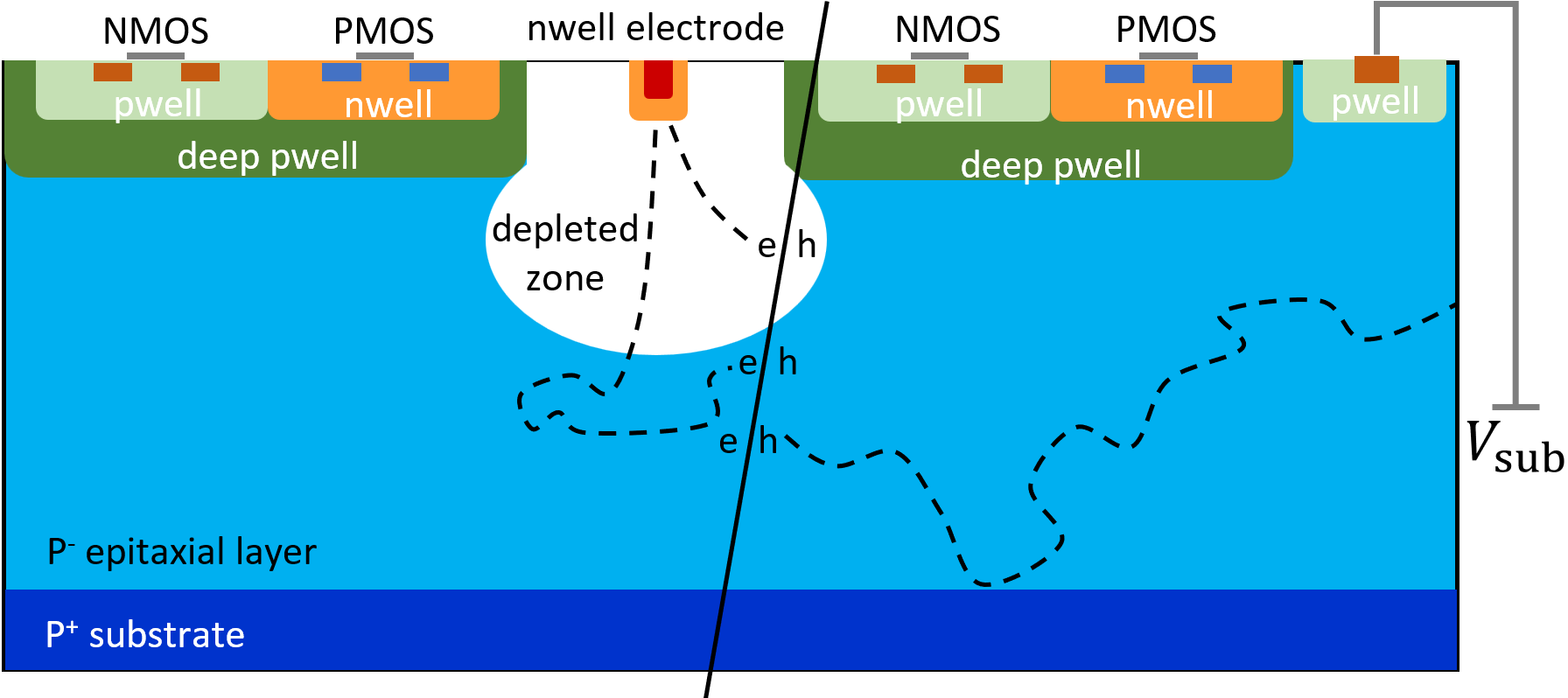}
\caption{Schematic diagram of the cross-section of the sensor when a reverse bias is applied to the P-substrate via a P-well.}
\label{fig:sensor}
\end{figure}

In the JadePix-3 sensor design, each pixel contains a P-well that allows for substrate biasing through a surface contact, as shown schematically in Fig.~\ref{fig:sensor}. This P-well is electrically contiguous with the underlying deep P-well and the P-type epitaxial layer. This configuration ensures that the entire P-type substrate is held at the applied potential, $V_{\rm sub}$. The pads connected to these P-wells are designed without ESD protection to allow for the application of a wide range of substrate bias voltages. When a reverse bias $V_{\rm sub}$ is applied on P-well and P-substrate, the depletion region around the N-well electrode expands. 

\subsection{Characterization Motivation}
Previous initial characterization proved that JadePix-3 had achieved the general design requirements under zero substrate reverse bias. The result revealed minimum thresholds ranging from 90 to 140 electrons, with a fake-hit rate below~$\rm 1 \times 10^{-10}$ per frame per pixel~\cite{jadepix3}. In the beam test for the JadePix-3 telescope, the sensor showed spatial resolutions of 5.2~$\rm{\upmu m}$ and 4.6~$\rm{\upmu m}$ in two dimensions, with a maximum detection efficiency approaching 99.0\% ~\cite{jadepix3telescope}. However, the full performance potential of the JadePix-3 sensor remains incompletely characterized, particularly concerning the effects of reverse bias.

\begin{figure}[h]
    \centering
    \includegraphics[width=0.7\textwidth]{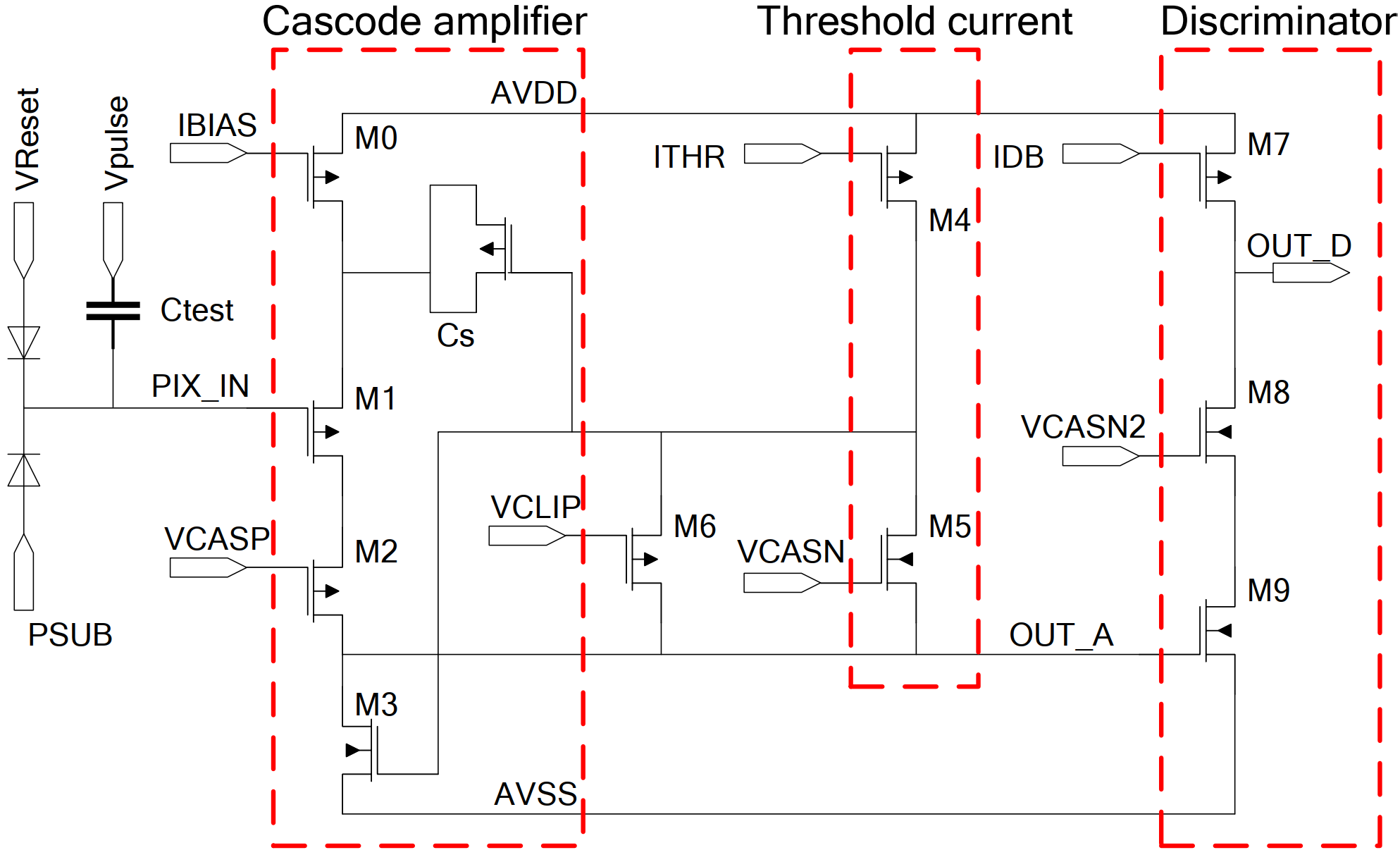}           
    \caption{Circuit topology for the JadePix-3 analog front-end~\cite{jadepix3}. This analog front-end was adapted from the ALPIDE~\cite{ALPIDE_upgrade} sensor, featuring low power consumption of 40~nW and minimum threshold below 100~$e^-$.}
    \label{fig:FE}
\end{figure}

The performance of MAPS is significantly influenced by the bias voltage applied to the substrate. This bias has a multifaceted impact: on one hand, it directly increases the size of the depletion zone and reduces the sensor's input capacitance~\cite{alpide}. On the other hand, it also affects the characteristics of the integrated circuitry; specifically, as the substrate bias becomes more pronounced, the threshold of NMOS transistors within the analog front-end circuitry (M3, M5, M8 and M9 in Fig.~\ref{fig:FE}) tends to increase due to the body effect, consequently altering the circuit's operating status. Therefore, adjusting the circuit's operating point after applying substrate bias and investigating the resulting impact on these key performance metrics becomes essential. These metrics are vital for the R\&D of the CEPC vertex detector.

Consequently, this work provides a thorough study of reverse bias effects, which is imperative for optimizing sensor design and ensuring its suitability for high-precision particle tracking applications.

This work also introduces an integrated and systematic test methodology (detailed in Section~\ref{sec:work-point}), which was applied to guide the subsequent lab-based characterization of MAPS behavior under varying reverse bias conditions.
Our approach systematically sweeps the reverse bias from 0 V to –6~V under identical threshold settings.
This methodology establishes comparable operating conditions across different bias voltages, thereby facilitating an unambiguous assessment of performance changes predominantly attributable to the bias variation itself.

\section{Experimental Methods} \label{sec:experimental setup}
\subsection{Measurement Setup} \label{sec:measurement setup}
\begin{figure}[h]
    \begin{minipage}[c]{0.235\textwidth}
    \centering
    \includegraphics[width=\textwidth]{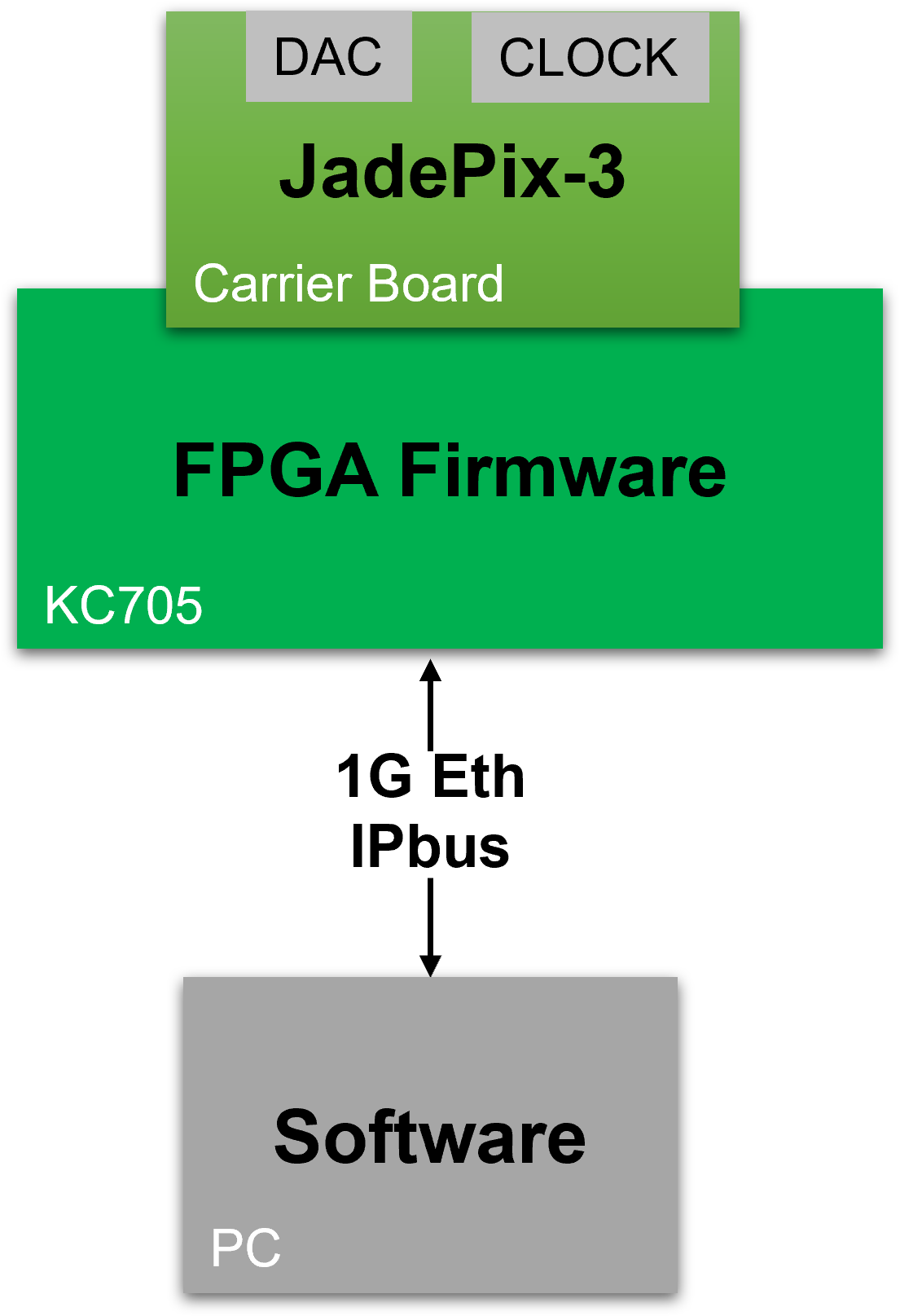}
    \subcaption{}
    \label{fig:setup_conceptual}
    \end{minipage} 
    \begin{minipage}[c]{0.755\textwidth}
    \centering
    \includegraphics[width=\textwidth]{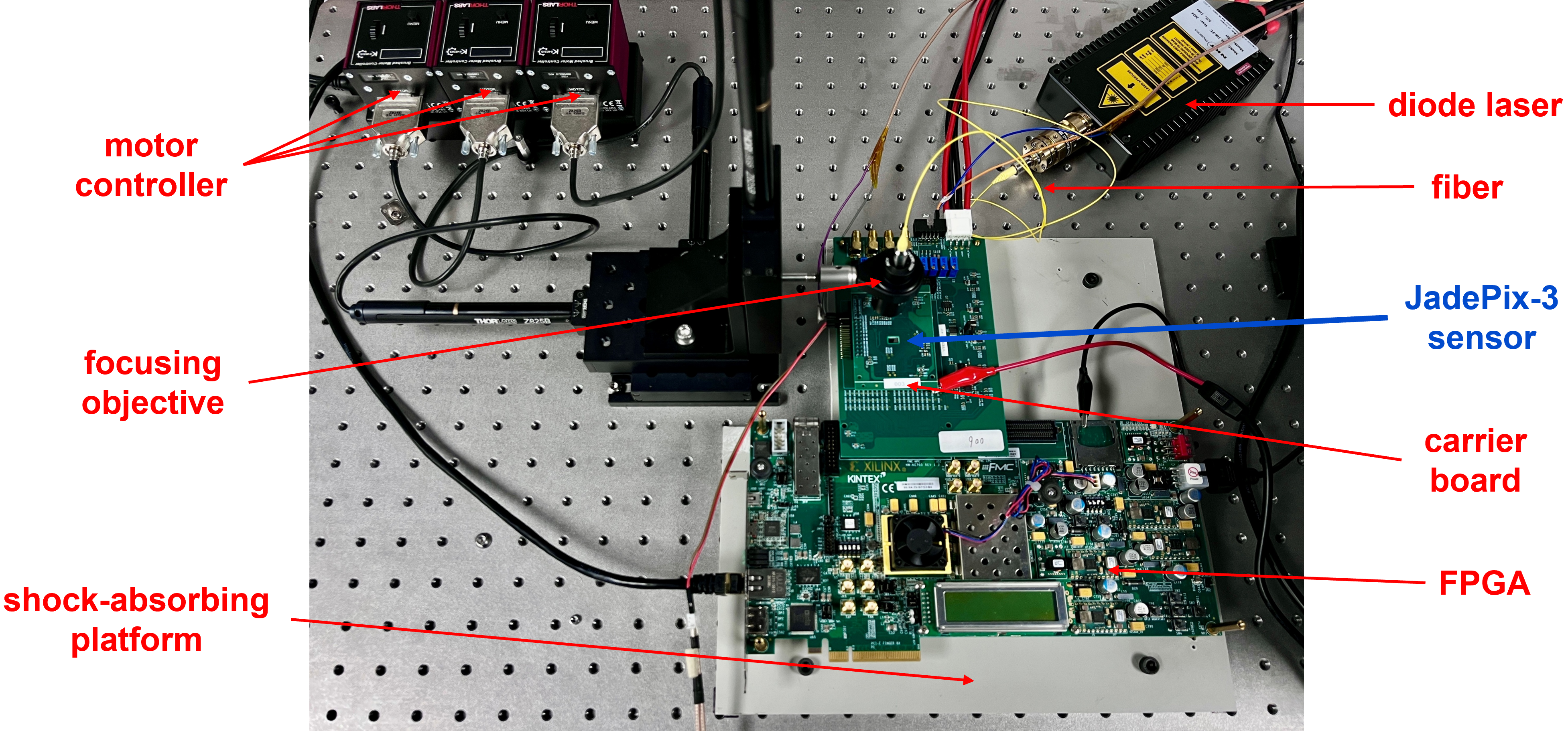}
    \subcaption{}     
    \label{fig:setup_laser}
    \end{minipage}
    \caption{(a) Schematic view of the test system; (b) Setup of the infrared laser test system. There is a window in the PCB to allow radiative source or laser beam incident to the JadePix-3 sensor. All devices are fixed or placed on an optical table.}
    \label{fig:setup}
\end{figure}

The experimental setup, which is based on the one described in Ref.~\cite{jadepix3}, is shown in Fig.~\ref{fig:setup}.
The JadePix-3 chip has been wire-bonded to a carrier board and then connected to an external FPGA-based DAQ system (see Fig.~\ref{fig:setup_conceptual}). The FPGA board configures the chip parameters and transfers data to the PC via Ethernet. A shock-absorbing platform was employed to maintain system stability and prevent displacement of the test source, either a radioactive source or a laser source, and the sensor during the measurements. A source meter is used to apply reverse bias ranging from 0 to -6~V to the substrate. The upper limit of -6~V was chosen as a conservative value to ensure a stable operation, based on breakdown voltages reported for other sensors fabricated in the same 180~nm process~\cite{alpide}.

To conduct laser experiment under the demand of high spatial resolution, a laser test system featuring three high-precision motor controllers is deployed (see Fig.~\ref{fig:setup_laser}). The spatial position in three orthogonal directions of the laser source is regulated precisely by these three motors, each has a repeatable position resolution of 1 $\rm{\upmu m}$ and a minimum step size of 0.1~$\rm{\upmu m}$. Details about the experiment parameters and test method will be discussed in Section~\ref{subsec:laser}.

\subsection{Working-point Selection}\label{sec:work-point}
This part details the methodology for selecting operational working-points of the JadePix-3 sensor. The primary objective is to establish uniform thresholds that enable the comparison of the sensor's performance across varying substrate reverse biases. Central to this process is the accurate measurement of pixel thresholds, achieved using the S-curve method.

The operational parameters of the pixels—threshold, fixed pattern noise (FPN), and temporal noise (TN)—were characterized using the standard ``S-curve'' method. This technique involves injecting a series of electrical pulses of increasing charge ($Q_{\rm injected}$) into each pixel and measuring the hit response. The injected charge is regulated by an external voltage pulse, $V_{\rm pulse}$, according to the relation $Q_{\rm injected} = V_{\rm pulse} \cdot C_{\rm test}$, where $C_{\rm test}$ is the injection capacitor with a designed value of 0.18~fF. The resulting S-shaped curve for each pixel is then analyzed to extract its specific threshold and noise characteristics. The final threshold value is converted to electrons via the value of $C_{\rm test}$.

\subsubsection{Target Working-Point Tuning}\label{sec:target}
To thoroughly assess the impact of reverse bias on sensor performance, it is essential to tune the sensor to \textit{target working-points} that exhibit a uniform average threshold across various substrate biases. Achieving such consistency poses a challenge due to the complex interplay of multiple parameters influencing the analog front-end. Therefore, the operating conditions—particularly the DAC settings that control the front-end bias condition—must be meticulously adjusted.

This work focuses on the characterization of sectors 0-2. Sectors 0-2 share the common DAC configuration. The analog front-end design is identical for sectors 0-2, with variations existing only in their digital logic blocks for pixel functions and in their respective pixel pitches (see Tab.~\ref{tab:digital}). Sector 3 is excluded from the present discussion; its front-end current was increased for radiation tolerance evaluations, leading to significantly higher power consumption.

The M6 MOSFET, biased by VCLIP, clamps excessively high OUT\_A signals (see Fig.~\ref{fig:FE}). Given that the amplitude of OUT\_A increases significantly with reverse bias due to a reduction in the sensor capacitance, VCLIP is set to 1.307~V. This is its maximum value and is chosen to eliminate the clamping effect of M6 on the OUT\_A amplitude during characterization. To compensate for the increase of the M8 gate threshold induced by substrate reverse bias, VCASN2 is increased from its nominal 400 mV to 722 mV, thereby maintaining a sufficient gate voltage for the cascode structure of M8 and M9.

VCASN and ITHR are the primary DAC settings in the JadePix-3 front-end that jointly determine the sensor's threshold. To ensure comparable conditions across different reverse bias settings ($V_{\rm sub} = 0, -1, \dots, -6\,$V), a threshold tuning step was performed by maintaining the ITHR value and adjusting the VCASN value for each $V_{\rm sub}$. The VCASN values were specifically chosen, as detailed in Tab.~\ref{tab:dac}, to achieve target thresholds around 205$e^-$ for sector 0, 225$e^-$ for sector 1 and 215$e^-$ for sector 2 across all bias levels. Since sectors 0-2 share a common DAC channel for both the VCASN and ITHR settings, the slightly different average thresholds are a consequence of applying a single, global DAC configuration to the three differently designed sectors. These DAC configurations define the seven \textit{target working-points} used for subsequent characterizations.

\begin{table}[h]
  \centering
  \caption{Nominal and Target operating configurations for the analog front-end. The seven Target working-points share constant ITHR, VCLIP, and VCASN2 settings, while VCASN is adjusted to achieve a similar average threshold across different $V_{\rm sub}$ levels.}
  \resizebox{\textwidth}{!}{
  \begin{tabular}{lrrrrr}
    \toprule
    Settings & $V_{\rm sub}$ & VCASN/mV & ITHR/pA & VCLIP/mV & VCASN2/mV \\
    \midrule
    Nominal & 0~V & 400 & 500 & 0 & 400 \\
    \cmidrule(lr){1-6}
    \multirow{7}{*}{\makecell[l]{Target}} & 0~V & 400 & \multirow{7}{*}{500} & \multirow{7}{*}{1307} & \multirow{7}{*}{722} \\
    & -1~V & 592 & & & \\
    & -2~V & 732 & & & \\
    & -3~V & 861 & & & \\
    & -4~V & 986 & & & \\
    & -5~V & 1129 & & & \\
    & -6~V & 1265 & & & \\
    \bottomrule
  \end{tabular}}
  \label{tab:dac}
\end{table}

The threshold distributions for these target working-points are presented in Fig.~\ref{fig:threshold}. It is observed that these distributions at non-zero reverse biases are noticeably broader compared to zero bias, indicating increased pixel-to-pixel threshold variations. This broadening of the threshold distributions is quantified in Fig.~\ref{fig:FPNvsBias}, which plots the FPN as a function of bias voltage. The corresponding TN analysis is shown in Fig.~\ref{fig:TNvsBias}. An increase in TN is observed under reverse bias conditions. The underlying reasons for this behavior will be discussed in Section~\ref{subsec:analogue} along with the change of the OUT\_A signal amplitude.

\begin{figure}[h]
    \centering
    \includegraphics[width=\linewidth]{fig/Threshold_Distributions_with_Nominal.png}
    \caption{Threshold distributions with average threshold tuned to 205$e^-$, 225$e^-$, 215$e^-$ in sector 0, 1, 2, respectively. The average and RMS values of each distribution are shown in the legend. The VCASN value is set respectively at the corresponding $V_{\rm sub}$ value as listed in Tab.~\ref{tab:dac}.}  
    \label{fig:threshold}
\end{figure}

\begin{figure}[h]
    \centering
    \begin{minipage}[c]{0.55\textwidth}
    \centering
    \includegraphics[width=\textwidth]{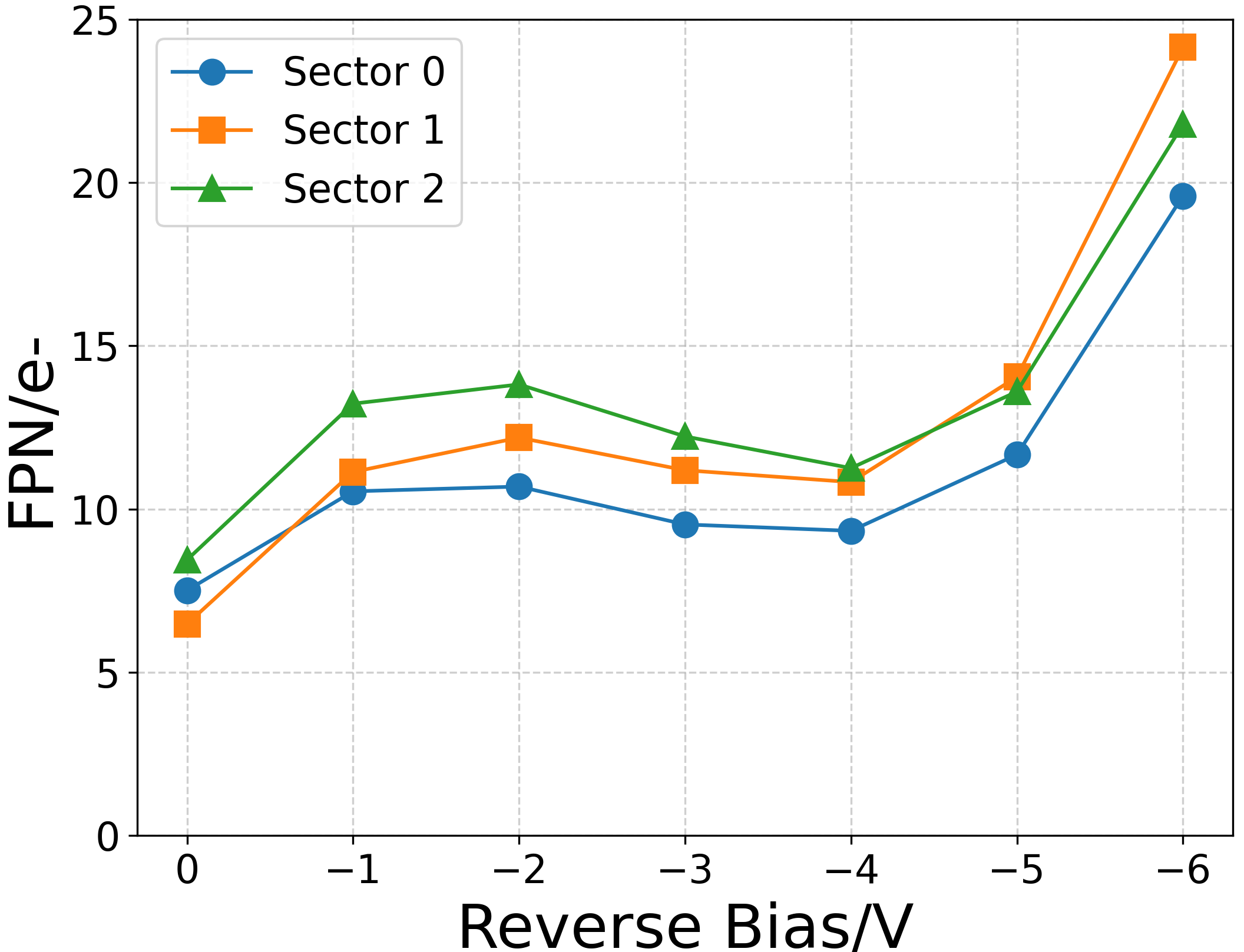}
    \subcaption{}     
    \label{fig:FPNvsBias}
    \end{minipage}
    \begin{minipage}[c]{0.55\textwidth}
    \centering
    \includegraphics[width=\textwidth]{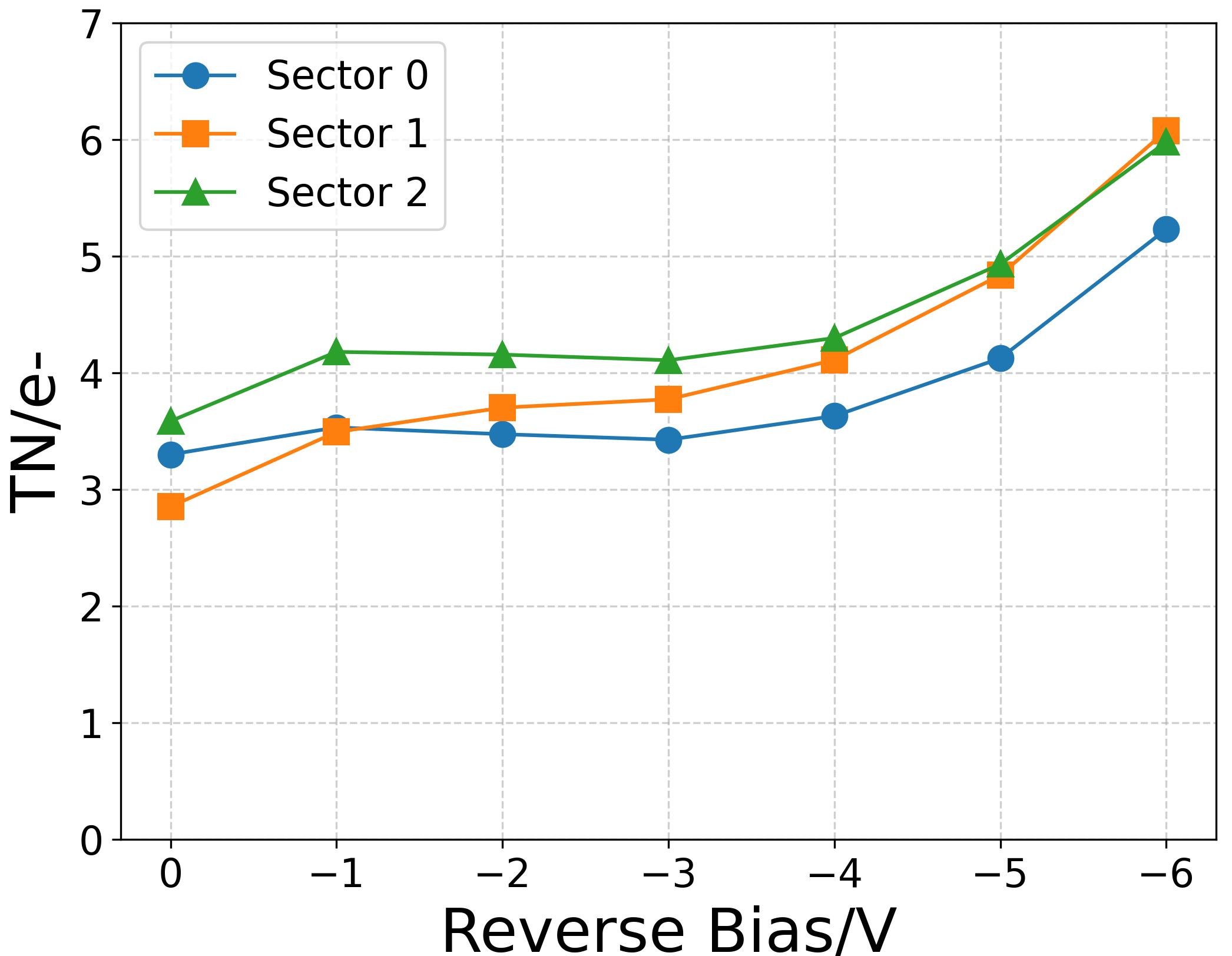}
    \subcaption{}     
    \label{fig:TNvsBias}
    \end{minipage}
    \caption{(a) Average FPN as a function of reverse bias. (b) Average TN as a function of reverse bias.}    
\end{figure}
\clearpage

\subsubsection{ITHR Scan}\label{sec:ithr_scan}
To investigate the sensor performance across a range of detection thresholds, the ITHR value was systematically scanned starting from each of these target working-points. This procedure yielded a series of operating conditions—7 distinct $V_{\rm sub}$ levels, each with 11 different ITHR settings—resulting in a set of \textit{working-points} with varying thresholds. 

\begin{figure}[h]
    \centering
    \begin{minipage}[c]{0.50\textwidth}
        \centering
        \includegraphics[width=\textwidth]{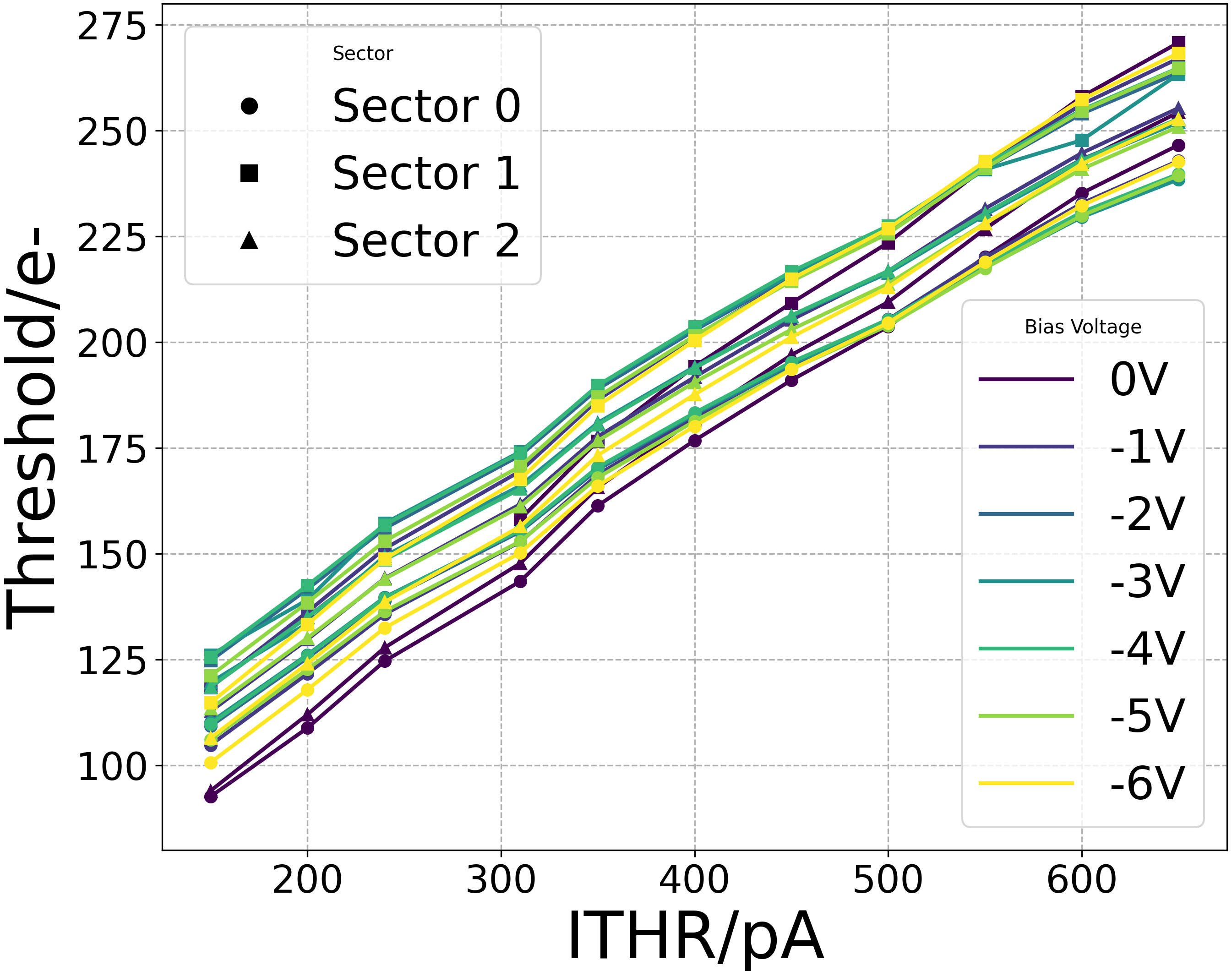}
        \subcaption{}
        \label{fig:ithrscanthreshold}
        \end{minipage} \\
    \begin{minipage}[c]{0.50\textwidth}
        \centering
        \includegraphics[width=\textwidth]{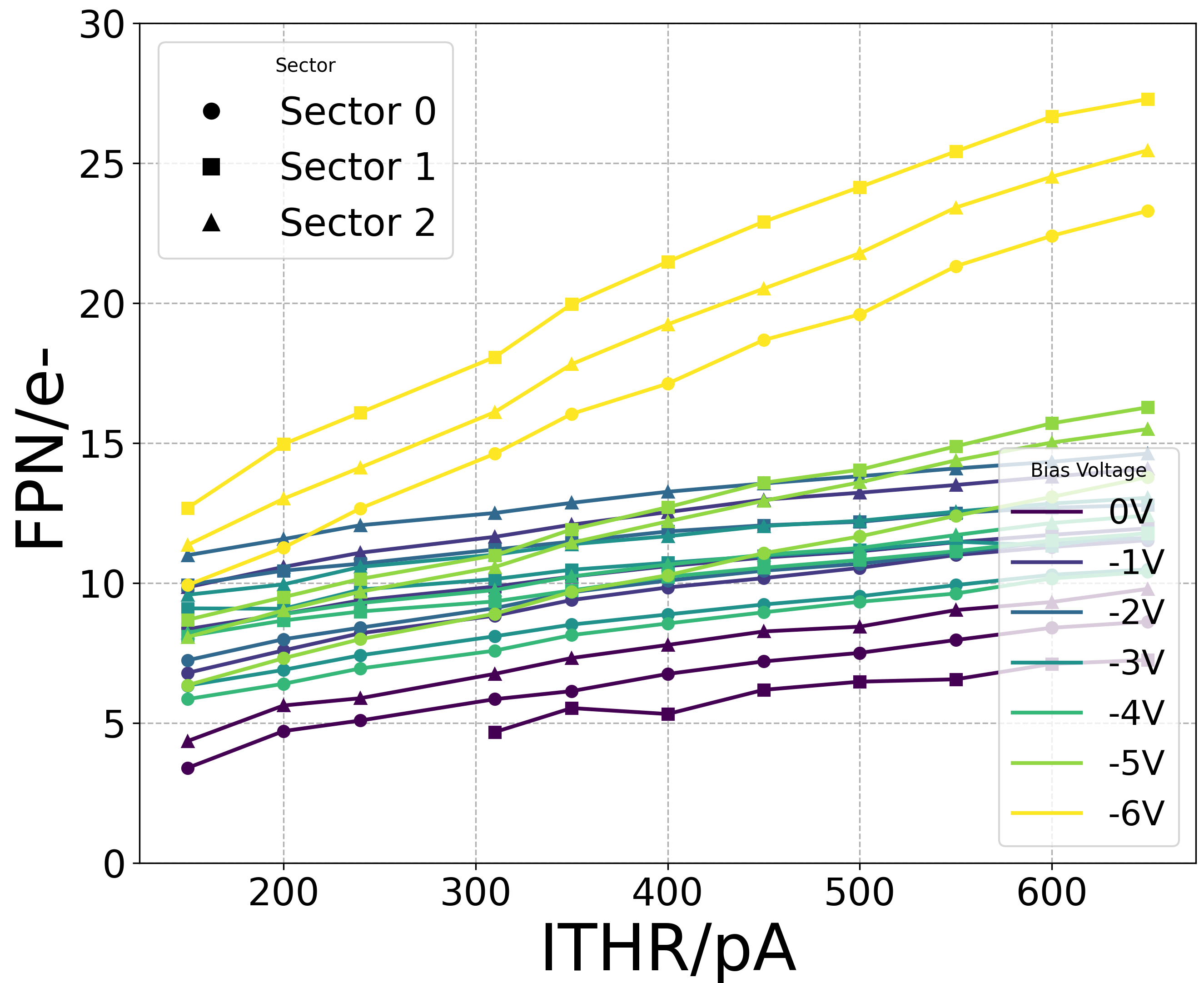}
        \subcaption{}    
        \label{fig:ithrscanfpn}
    \end{minipage}\\
    \begin{minipage}[c]{0.50\textwidth}
        \centering
        \includegraphics[width=\textwidth]{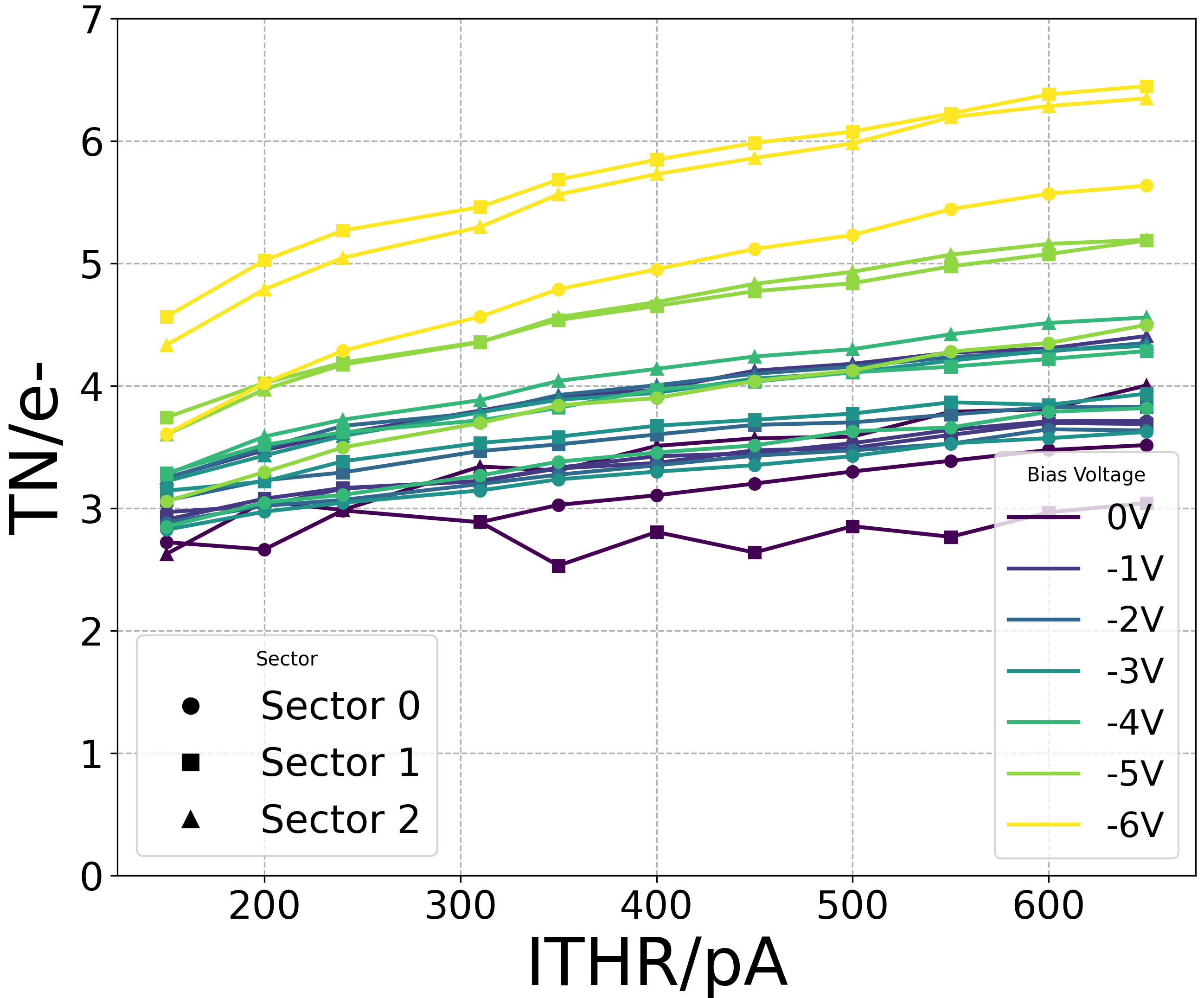}
        \subcaption{}     
        \label{fig:ithrscantn}
    \end{minipage}    
    \caption{ITHR scan results. (a) Threshold, (b) FPN (fixed pattern noise), and (c) TN (temporal noise) versus ITHR value for working-points of distinct reverse biases.}
        \label{fig:ithrscan}
\end{figure}

Fig.~\ref{fig:ithrscan} illustrates the impact of the ITHR value on the mean threshold, FPN and TN, respectively. Both FPN and TN generally exhibit an increasing trend with higher threshold values.

The linear relationship between the mean threshold and the ITHR setting, depicted in Fig.~\ref{fig:ithrscanthreshold}, is consistently observed across all tested reverse bias voltages. This linearity not only confirms the consistency of the threshold tuning but also underscores that ITHR is a suitable parameter for threshold control under different reverse biases.

\clearpage

\section{Experimental Results and Discussion}\label{sec:result}
 Reducing the sensor capacitance and expanding the depletion zone are the most intrinsic and direct consequences of applying a reverse bias on the substrate. After adjusting thresholds under different reverse biases and determining a set of working-points by the ITHR value scan described in Sec.~\ref{sec:ithr_scan}, a thorough sensor characterization regarding the impact of the substrate reverse bias voltage was conducted, including leakage current, analog response, fake-hit rate and charge collection efficiency.
 
\subsection{Leakage Current}\label{subsec:leakage current}
In the absence of ionizing radiation, thermally generated free charge carriers are collected by the diode, resulting in what is known as \textit{leakage current} or \textit{dark current}. Leakage current from the pixel area impacts the sensor's electrical performance by inducing shot noise~\cite{shotnoise}.

In the JadePix-3 pixel architecture (Fig.~\ref{fig:FE}), the pixel's leakage current, under quiescent conditions, flows from the collection electrode (N-well) to the P-substrate, causing a gradual discharge of the collection electrode. 
The current supplied through the VReset pin, which is biased by an external DAC and applied to the collection electrode via a forward-biased diode, is used to restore the collection electrode's potential, and its magnitude is therefore equal to this leakage current. Consequently, the total leakage current from sectors 0-2 was measured using a source meter connected to the VReset readout pin. The measurement was carried out at a room temperature of 25$^\circ {\rm C}$ in darkness to suppress the photocurrent\footnote{Initial measurements conducted with this setup but without rigorous light shielding showed a current of approximately 1.2~nA. This value, predominantly attributed to photocurrent, is more than an order of magnitude higher than the leakage current values reported herein, underscoring the critical importance of dark condition for accurate leakage current characterization.}. 

\begin{figure}[h]
    \centering
    \includegraphics[width=0.55\textwidth]{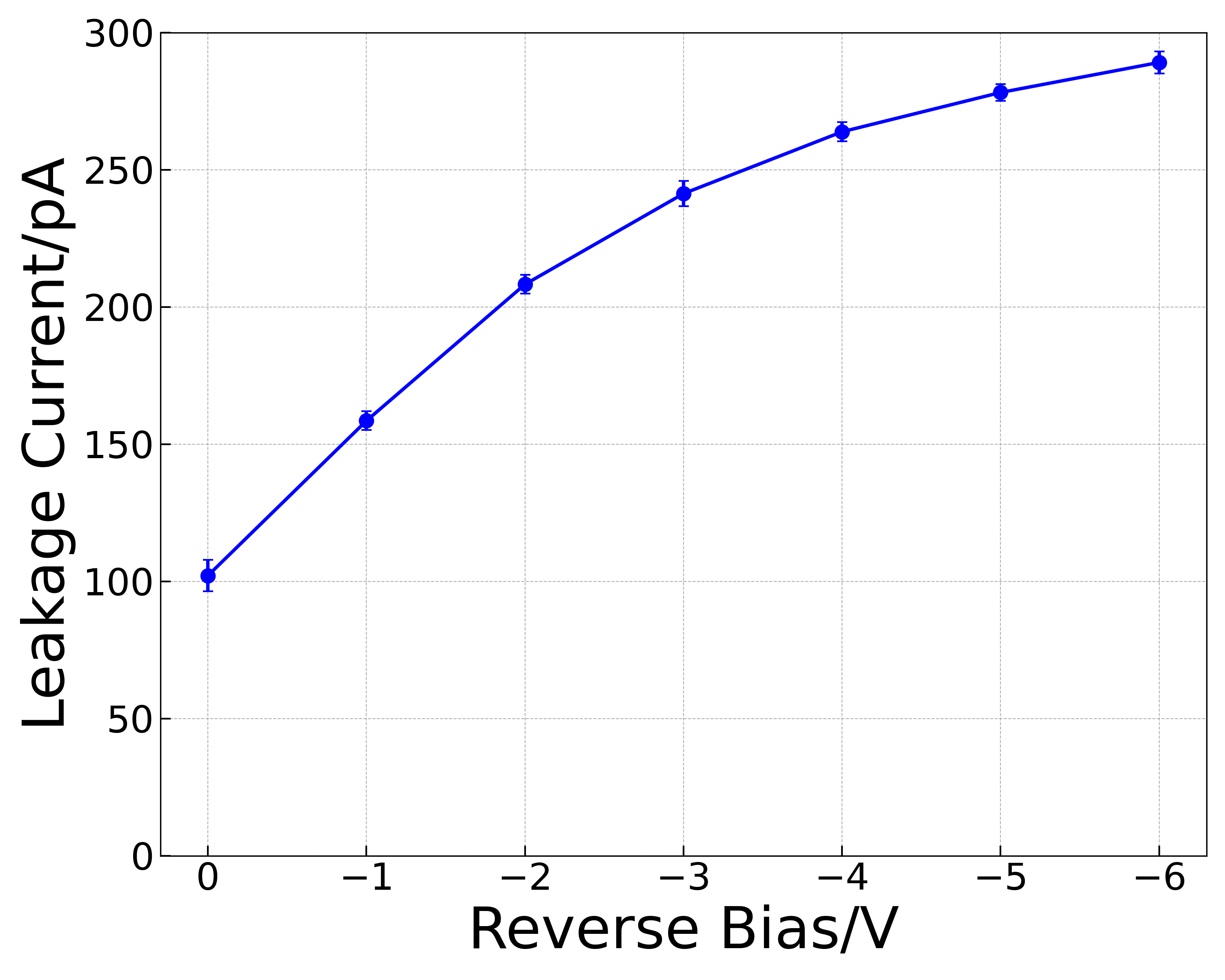}
    \caption{Total leakage current from sectors 0-2 (encompassing 73,728 pixels) as a function of reverse bias, measured in darkness at 25$^\circ {\rm C}$.}
    \label{fig:leakage}
\end{figure}

The results, presented in Fig.~\ref{fig:leakage}, indicate that the leakage current from the pixel area is small (less than 3~fA per pixel) and increases with the reverse bias applied to the substrate. The current observed at substrate bias 0V is a result of the potential difference between the collection node, which is held at +1V by the VReset pin, and the grounded substrate. This observed increase with reverse bias is consistent with the understanding that a primary component of leakage current originates from the thermal generation of electron-hole pairs within the depletion region of the pixel diode. As the substrate reverse bias increases, the depletion region expands, thereby enlarging the volume for carrier generation and consequently leading to a higher measured leakage current~\cite{pixeldetectors}.

\subsection{Input Capacitance}
\label{subsec:analogue}
The effective pixel input capacitance of the sensor, $C_{\rm effective}$, can be determined by analyzing the analog response of the sensor under different reverse bias voltages. For test purposes, the chip is equipped with several dedicated test pixels. The analog signal OUT\_A from one of these selectable pixels can be connected to an external pin, allowing it to be monitored with an oscilloscope. The analog output (denoted as OUT\_A in Fig.~\ref{fig:FE}) amplitude is given by:
\begin{equation}
    A_{\rm{OUT\_A}}=\frac{Q_{\rm injected}}{C_{\rm effective}} \cdot f_{\rm gain} \cdot f_{\rm sf}~,\label{eq:amp}
\end{equation}
where $A_{\mathrm{OUT\_A}}$ is the amplitude of OUT\_A, $Q_{\rm injected}$ is the injected charge which equals to $C_{\rm test} \cdot V_{\rm pulse}$ in the electric pulse test, $f_{\rm gain}$ is the gain factor from the front-end amplifier, and $f_{\rm sf}$ is the factor of the source follower as a buffer from the test pixel. $f_{\rm gain}$ varies with signal amplitude because the front-end is a non-linear amplifier. For the estimation of $C_{\rm effective}$, these factors were set to $f_{\rm sf}=0.7, f_{\rm gain}=20$, which are derived from post-layout SPICE simulations of the front-end circuit design.

Two independent and complementary methods were employed in this study to measure $C_{\rm effective}$ via the relation:
\begin{equation}
    C_{\rm effective}=C_{\rm test} \cdot \frac{V_{\rm pulse}}{A_{\rm {OUT\_A}}} \cdot f_{\rm gain} \cdot f_{\rm sf}~.\label{eq:ceffective}
\end{equation}
\vspace{-6ex}
\begin{itemize}
    \item \textbf{Method 1}: In this approach, the pulse voltage was fixed at $V_{\rm pulse}=160 {\rm mV}$ while the analog response amplitude was recorded via an oscilloscope. The measurement was repeated to extract a statistically robust value for the amplitude. The measured amplitude is then substituted into Eq.~(\ref{eq:ceffective}), the term $A_{\rm{OUT\_A}}$, to obtain $C_{\rm effective}$.
    \item \textbf{Method 2}: In this method, the analog response amplitude, $A_{\rm{OUT\_A}}$, was maintained at a constant value by dynamically adjusting $V_{\rm pulse}$. The corresponding waveform measurement was repeated, and the adjusted $V_{\rm pulse}$ value (required to hold the amplitude constant) was then used in Eq.~(\ref{eq:ceffective}) to compute $C_{\rm effective}$.
\end{itemize}

\begin{figure}[ht]
    \centering
    \begin{minipage}[c]{0.495\textwidth}
        \centering
        \includegraphics[width=\textwidth]{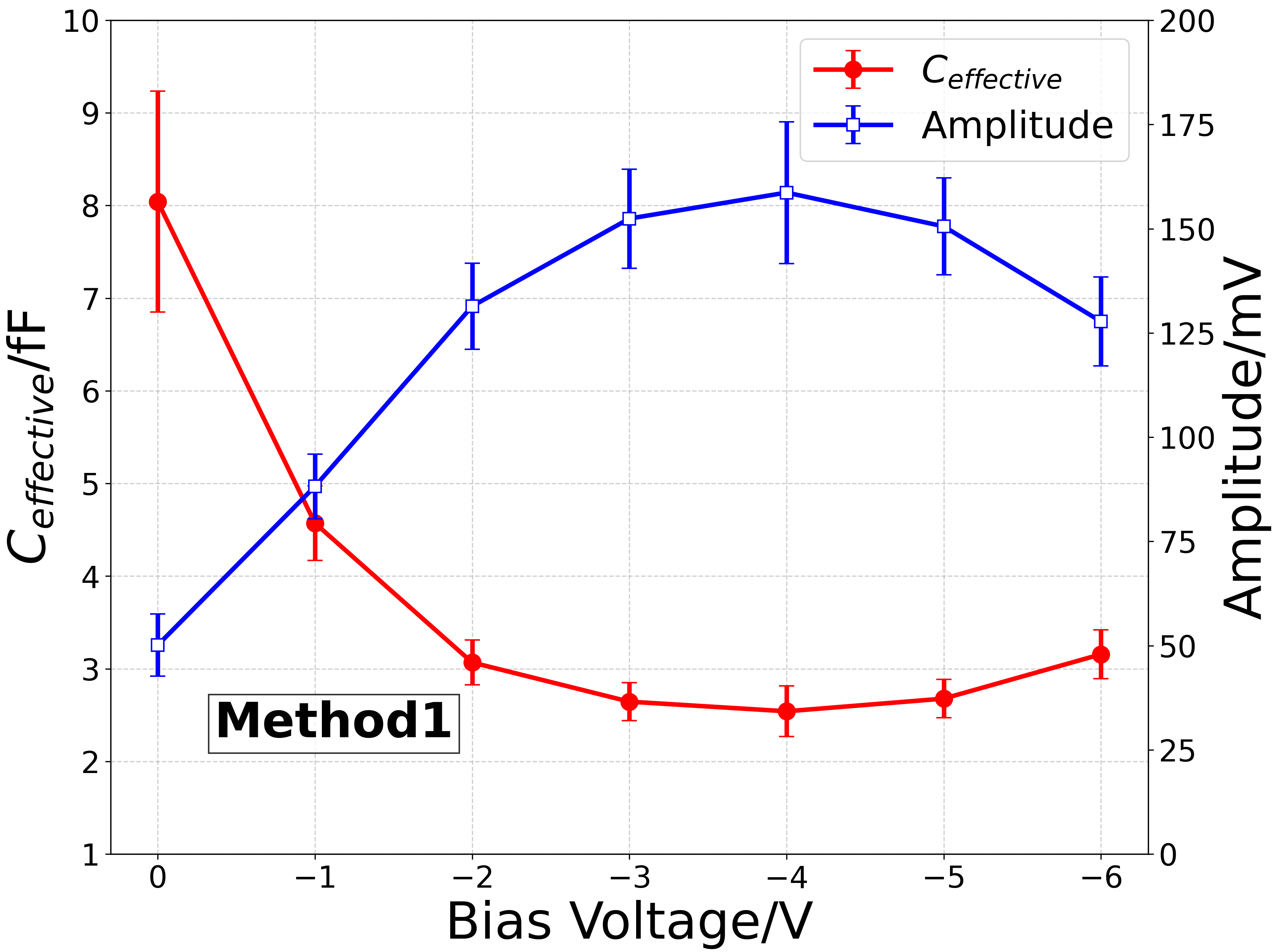}
        \subcaption{}
        \label{fig:method1}
        \end{minipage}
    \begin{minipage}[c]{0.495\textwidth}
        \centering
        \includegraphics[width=\textwidth]{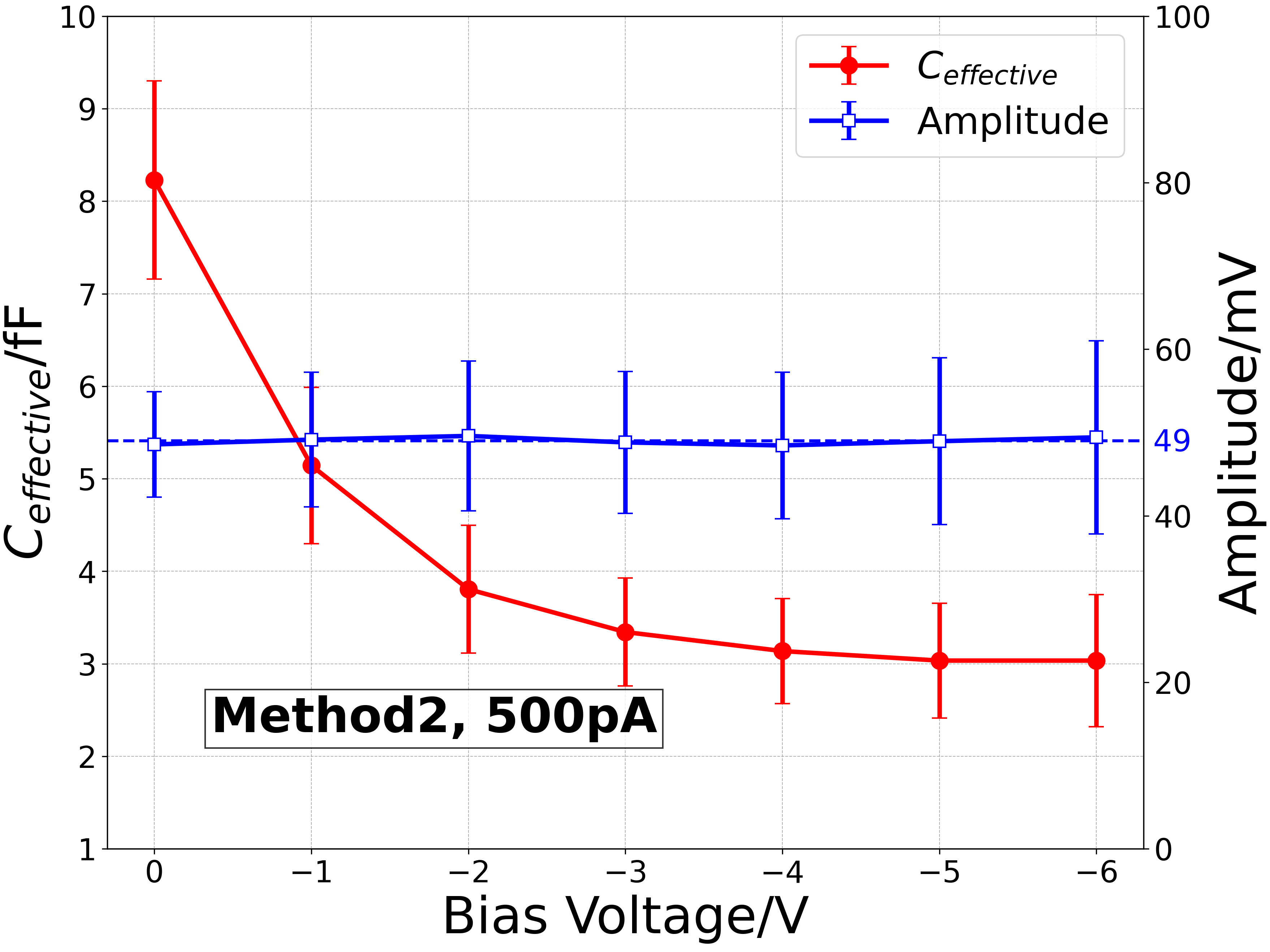}
        \subcaption{}    
        \label{fig:method2}
    \end{minipage}
    \caption{Effective capacitance measured as a function of the reverse bias voltage using two methods. (a)\textbf{Method 1:} $V_{\rm pulse}$ is fixed at 160~mV. (b)\textbf{Method 2:} The analog amplitude is fixed at 49~mV by adjusting $V_{\rm pulse}$ accordingly. For both methods, the calculation of $C_{\rm effective}$ assumes $f_{\rm gain}=20$ and $f_{\rm sf}=0.7$ and ITHR is set at 500~pA (target working-point condition). The amplitude shown is the mean value from 200 repeated waveform acquisitions; the error bars represent the standard deviation (RMS) of the distribution, corresponding to the measured electronic noise.}
    \label{fig:analog result}
\end{figure}

The two methods provide complementary insights. While Method 1 directly captures the amplitude response at a fixed pulse voltage, Method 2, by contrast, keeps the amplitude-dependent $f_{\rm gain}$ factor fixed via maintaining the amplitude value constant, which offers a more reliable $C_{\rm effective}$ result with increasing reverse bias voltage.

Results for these two methods after calculation are shown in Fig.~\ref{fig:analog result}. In both methods, the input capacitance can be reduced by more than 50\% when going from $V_{\rm sub}=0~{\rm V}$ to $V_{\rm sub}=-2~{\rm V}$, corresponding to the rapid expansion of the depletion region. This reduction is very important as smaller input capacitance increases the charge-voltage conversion coefficient at the input of the front-end circuit for a given charge. As a consequence of the increased charge-voltage conversion, the amplitude of the analog signal (OUT\_A) increases significantly with reverse bias.  
In Section~\ref{sec:target}, to keep the input-referred threshold constant, the VCASN was tuned at different reverse bias voltage. The tuning of VCASN actually lowers the effective OUT\_A baseline which is relative to the voltage required to switch the transistor M9 on. The increase of FPN and TN in Fig.~\ref{fig:FPNvsBias} and \ref{fig:TNvsBias}, respectively, may be related to the tuning of VCASN because the same effect observed in a dedicated VCASN scan shown in Ref.~\cite{jadepix3} (Fig.~13).
\clearpage

\subsection{Fake-hit Rate Test}\label{subsec:fhr}
Fake-hit, i.e., spurious signals that do not correspond to any external particle interactions, are a critical parameter in pixel detector performance. These false signals often arise from electronic noise, crosstalk, or other sensor imperfections, leading to erroneous track reconstruction. Therefore, assessing and reducing the fake-hit rate is crucial for ensuring the reliability of tracking detectors.

JadePix-3 employs both rolling shutter (RS) and global shutter (GS) readout sequence modes. In the conventional rolling shutter mode, one row is activated for reading at a time. This process requires 192~ns to transmit hit data and reset the pixels, and scanning the entire 512-row matrix takes 98.3~$\rm{\upmu s}$. While in global shutter, all 512 rows are frozen for readout after a period of acquisition time. To compare the two readout sequence modes, the acquisition time of the GS mode is set equivalent to 98.3~$\rm{\upmu s}$ and the fake-hit rate for each sector \( i \) in both modes can be calculated using a unified formula:

\begin{equation} \label{eq:fhr}
    R_i^{\text{mode}} = \frac{N_i^{\text{mode}}}{N_{\text{frames}}^{\text{mode}} \cdot N_{\text{pixels}}},
\end{equation}

where \( N_i^{\text{mode}} \) represents the total number of false hits in sector \( i \), \( N_{\text{frames}}^{\text{mode}} \) is the number of frames collected, and \( N_{\text{pixels}} = 512 \times 48 \) is the number of pixels per sector. For RS mode, the test system can readout continuously while for the GS mode the test system has to launch one readout for one frame, which limits the total number of frames that can be collected in the test. In the absence of external sources, a total of 100,000 frames (\( N_{\text{frames}}^{\text{RS}} = 100,000 \)) are collected in the RS mode, while 1,000 frames (\( N_{\text{frames}}^{\text{GS}} = 1,000 \)) are collected in the GS mode.

The sensitivity limit is defined as the lowest measurable fake-hit rate in the measurement and is determined by the number of frames for these two modes. Based on Eq.~\ref{eq:fhr}, this results in a sensitivity limit of $\rm 4\times 10^{-10}$ for the RS mode and $\rm 4\times 10^{-8}$ for the GS mode.

As the fake-hit rate is expected to be strongly influenced by the threshold, a threshold scan under different reverse biases was performed. 
The results are shown in Fig.~\ref{fig:fakehit}, which compares the fake-hit rate under two conditions: with all pixels included and with hot pixels masked. For the masked dataset, hot pixels were identified as those that contribute most to the false hits; specifically, up to 10 pixels (accounting for 0.04\%) in each sector with the highest fake-hit counts were masked.

According to the result, the difference between the two modes is insignificant above the GS sensitivity limit. Reverse bias substantially reduces the fake-hit rate compared with the unbiased state, with $V_{\rm sub}=-3~{\rm V}, -4~{\rm V}$ yielding the lowest values.

\begin{figure}[ht]
    \centering
    \includegraphics[width=\textwidth]{fig/FHR_Combined_Mask_NoMask.png}
    \caption{
    Fake-hit rate as a function of threshold measured with RS (rolling shutter) and GS (global shutter) modes under various reverse biases. The plot compares results with all pixels included versus those with hot pixels masked. For the masked data, up to 10 pixels per sector (0.04\%), corresponding to those with the highest fake-hit counts, were masked. The sensitivity limits of the measurement are indicated by the horizontal lines in each pad.}
    \label{fig:fakehit}
\end{figure}

\subsection{Radioactive Source Test}\label{subsec:radio}
Two radioactive sources, $^{90} \rm Sr$ and $^{55} \rm Fe$, were employed to assess the sensor's response to external ionizing radiation. 
$^{90} \rm Sr$, a $\beta$ emitter, generates electrons that traverse the sensor and deposit charge over an extended region. This distributed ionization is particularly sensitive to changes in the sensor’s depletion characteristics and charge-sharing behavior under reverse bias, thereby providing insights into the sensor’s charge collection efficiency. 
$^{55} \rm Fe$ emits 5.9~keV X-rays that interact with silicon primarily via the photoelectric effect and deposit an ionization charge of 1640 $e^-$ in a small point-like spatial region compared to the dimensions of the sensor, resulting in highly localized charge deposition. This characteristic makes $^{55} \rm Fe$ an ideal probe for evaluating the sensor’s response to concentrated ionization, providing a signal whose magnitude is comparable to that of a minimum ionizing particle (about 66 electron-hole pairs are generated per micron~\cite{bichsel1988straggling} in the 20~$\rm{\upmu m}$ epitaxial layer of the JadePix-3 prototype).

\begin{figure}[h]
    \centering
    \includegraphics[width=0.6\textwidth]{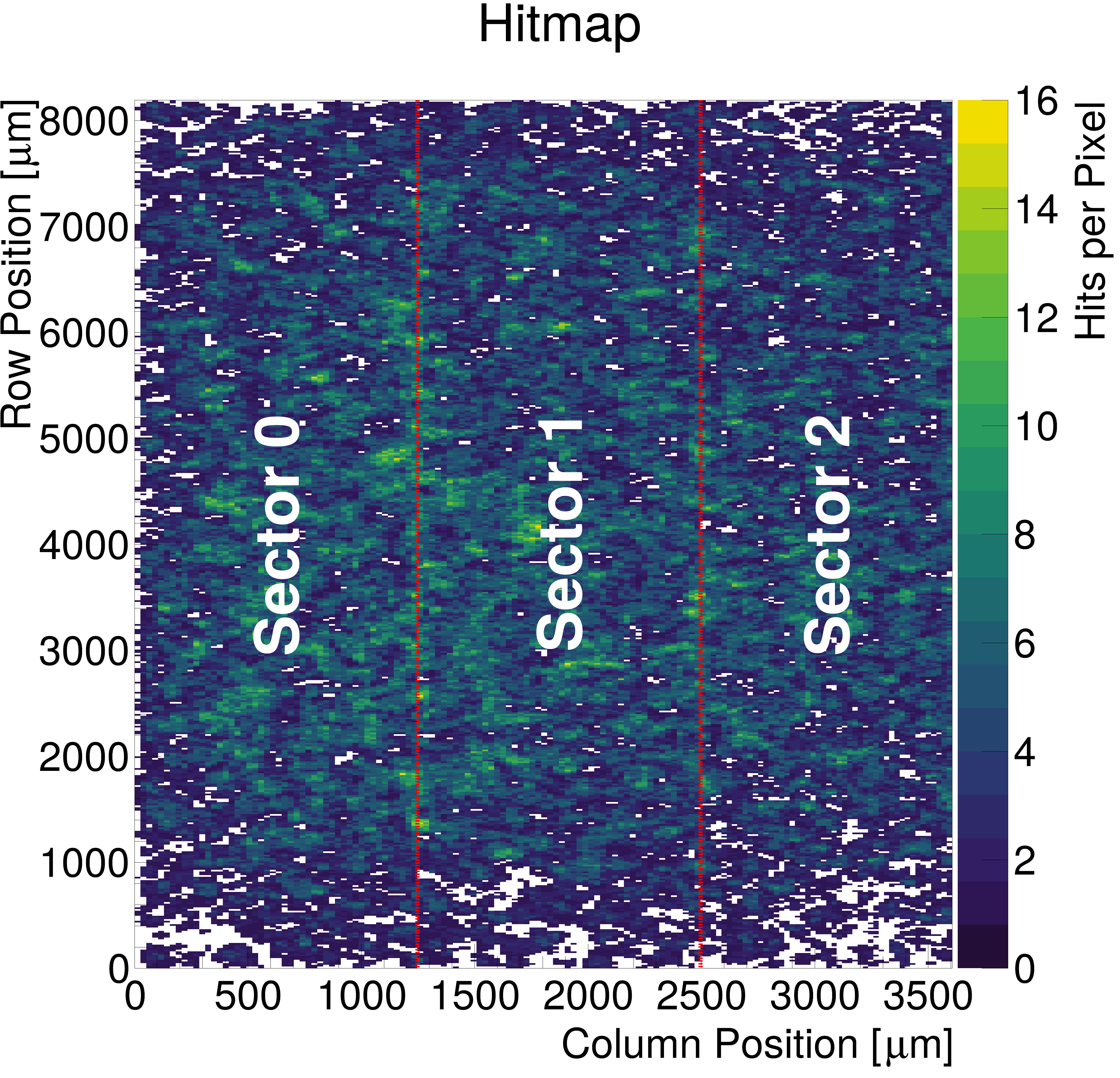}
    \caption{Hitmap of sectors 0,1,2 from the $^{90}$Sr source test, showing the number of hits per pixel. A total of 27,414 clusters (8,969 in Sec. 0, 10,053 in Sec. 1, and 8,392 in Sec. 2) were recorded under $\beta$-ray irradiation from the backside during a time duration of 50~s, with ITHR value fixed at 500~pA and $V_{\rm sub}$ at 0~V. The pixels at the edge of the array were partially obstructed by the PCB.}
    \label{fig:hitmap}
\end{figure}

For the source tests, an uncollimated $^{90} \rm Sr$ source was placed 2~mm above the sensor, and a $^{55} \rm Fe$ source was positioned 8~mm above it. For both, data were collected with a 100,000-frame rolling shutter scan. Due to charge sharing, the signal from a single particle interaction is often collected by several neighboring pixels, which form a cluster. The cluster size is defined as the number of pixels in such a group. Approximately 30,000 clusters were recorded in sector 0,1,2 over the course of 50 seconds, for which the hitmap from $^{90} \rm Sr$ source test is shown in Fig.~\ref{fig:hitmap}, where the pixels at the edge of the array were blocked partially by the PCB. 

A \textit{double hit} is defined as an event where the same pixel registers a hit in two consecutive frames. In this untriggered measurement, to ensure each particle hit is counted only once, the analysis retains the first hit from such an event but excludes the subsequent hit in the consecutive frame.

\begin{figure}[h]
	\centering
	\begin{minipage}[c]{0.6\textwidth}
		\centering
		\includegraphics[width=\textwidth]{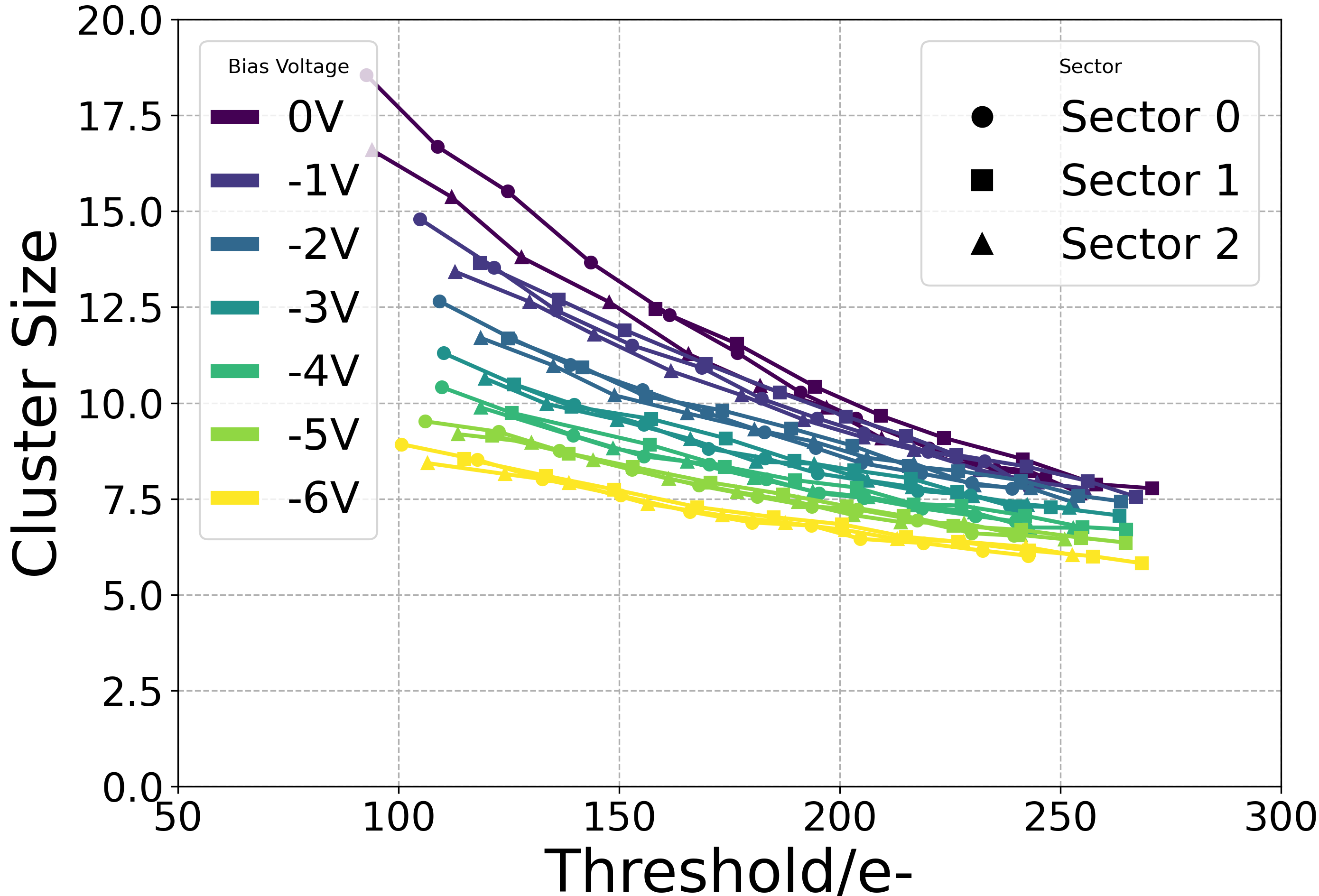}
		\subcaption{}
		\label{fig:SourceCS_Sr}
	\end{minipage}\\
	\begin{minipage}[c]{0.6\textwidth}
		\centering
		\includegraphics[width=\textwidth]{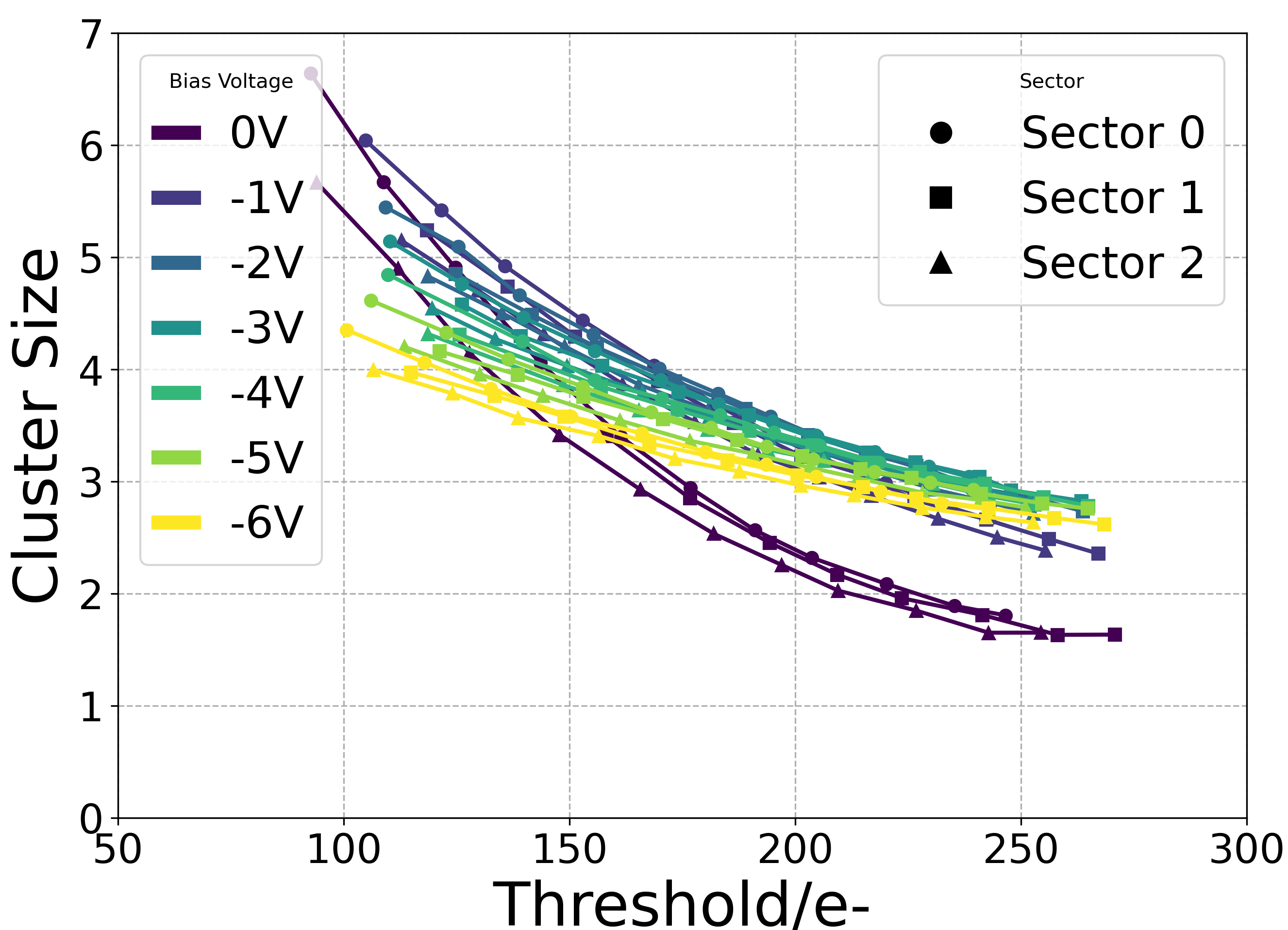}
		\subcaption{}
		\label{fig:SourceCS_Fe}
	\end{minipage}
	\caption{Average cluster size as a function of threshold for different reverse bias voltages, measured with (a) the $^{90} \rm Sr$ source and (b) the $^{55} \rm Fe$ source.}
	\label{fig:SourceCS}
\end{figure}

The results of cluster size analysis for both the $^{90}$Sr and $^{55}$Fe sources are presented in Fig.~\ref{fig:SourceCS}, which plots the average cluster size as a function of threshold for different bias voltages. It should be noted that the average cluster size is calculated using only detected events; particles that do not produce a signal above threshold are not included in this average.

Fig.~\ref{fig:SourceCS_Sr} demonstrates a clear reduction in cluster size with increasing reverse bias. At a threshold of 100~$e^-$, for instance, the average cluster size is reduced from 16-17 pixels at 0~V bias to 8-9 pixels at -6~V bias.
This reduction is expected considering that the stronger electric field from the applied reverse bias lowers the lateral diffusion of charge carriers. Consequently, more charge from particle interactions is collected locally within the seed pixel, reducing the spread to its neighbors. As a result, fewer surrounding pixels exceed the detection threshold, leading to a smaller average cluster size. It is worth noting that a \textit{certain amount} of charge sharing is often desirable as it improves spatial resolution~\cite{alpide}, which is important for the CEPC application.

The slope of cluster size as a function of the threshold also decreases significantly in both Fig.~\ref{fig:SourceCS_Sr} and Fig.~\ref{fig:SourceCS_Fe}. This is an indicator of improved charge collection efficiency. A steep slope (as seen at 0~V) signifies that the measured cluster size is highly sensitive to the threshold setting, which is characteristic of charge being spread by diffusion over many pixels. Conversely, a flatter slope indicates that the cluster size is less dependent on the threshold. This is a signature of efficient charge collection, where a strong electric field concentrates the majority of the charge into the seed pixel, making the resulting cluster size insensitive to threshold variations.

\clearpage

\subsection{Infrared Laser Test}
\label{subsec:laser}
Following the configuration of the laser test platform outlined in Section~\ref{sec:measurement setup}, a systematic investigation of the in-pixel laser hit efficiency was performed using an infrared laser beam incident from the backside of the chip. The laser operates at a wavelength of 1064~nm. The laser’s pulse energy is adjustable between 0 and 60~pJ, modulated via an attenuation factor ranging from 0\% tune (maximum pulse energy) to 100\% tune (minimum pulse energy), controlled through the Laser Power Tune function of the system. For this laser measurement, the selected pixel is located in sector~2, which has the smallest pixel pitch among the sectors (see Tab.~\ref{tab:digital}).

\subsubsection{Beam Position Adjustment}
To ensure optimal focusing and achieve the minimum laser spot size, which can be tuned to 3.4~$\rm{\upmu m}$, significantly smaller than the pixel size, at the depth of the chip’s sensitive layer, a scan was performed along the Z-direction (perpendicular to the sensor plane) to locate the beam waist, a method previously employed in the characterization of this sensor~\cite{jadepix3}.

\begin{figure}[h]
    \centering
    \includegraphics[width=0.6\textwidth]{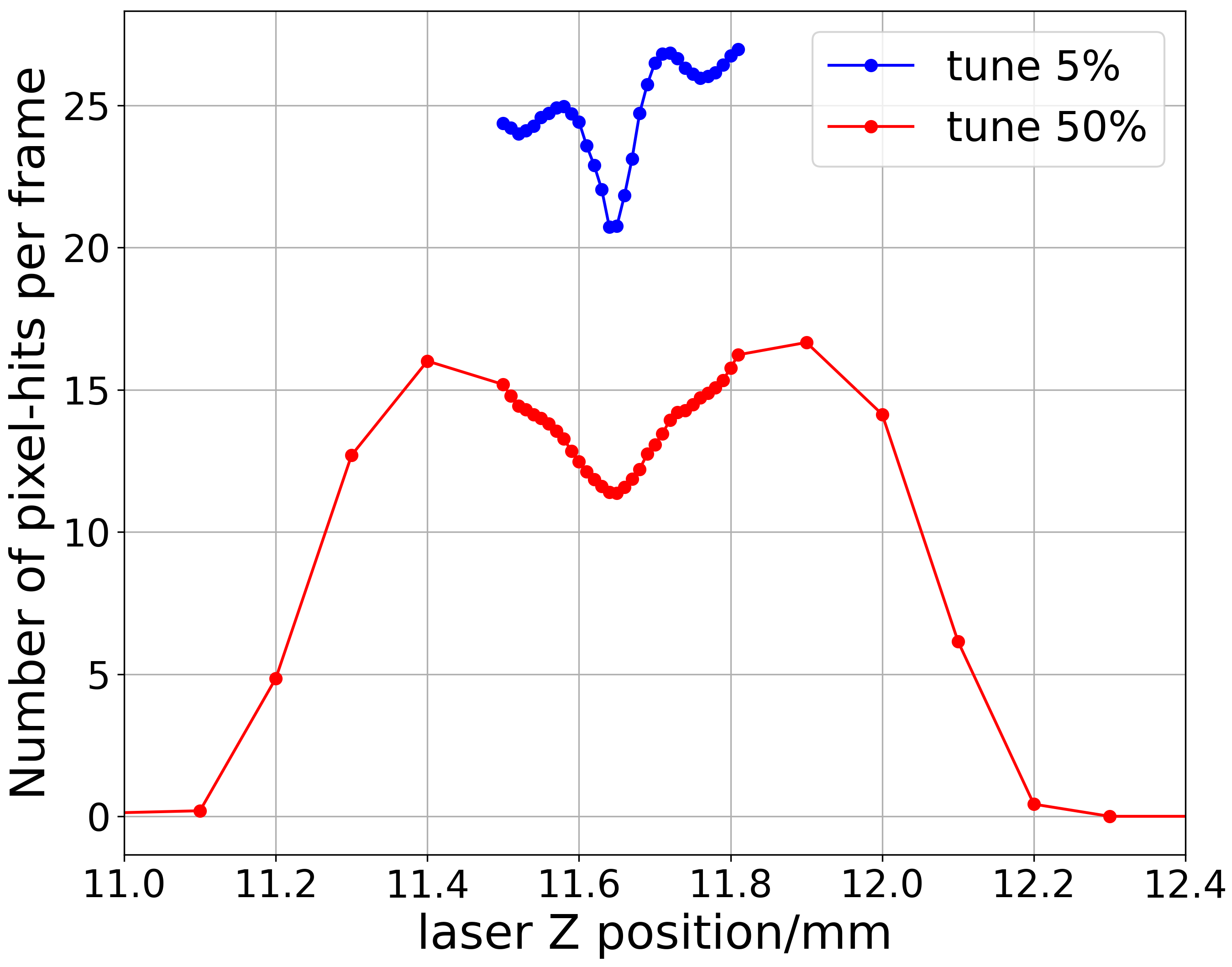}
    \caption{Average number of pixel hits per frame as a function of the laser position Z. The upper and lower curves are for the laser power attenuation at 5\% and 50\%, respectively. Settings of this experiment: laser frequency = 25~kHz, sensor threshold = 215~$e^-$, $V_{\rm sub}=0{\rm V}$, Rolling Shutter frame number = 1,000.}
    \label{fig:zscan}     
\end{figure}

This scan involves measuring the average number of pixel hits per frame over an acquisition of 1,000 frames as a function of the laser’s Z-position, which indicates the vertical distance to the sensor surface. The scan was performed at two attenuation levels of the laser: 5\% and 50\%. As shown in Fig.~\ref{fig:zscan}, the minimum number of hits, indicative of the beam waist, was observed at a Z-coordinate of 11.65~mm. This position was subsequently fixed for all subsequent measurements to ensure optimal focus and minimal spot size on the chip.

\begin{figure}[h]
    \centering
    \includegraphics[width=0.9\textwidth]{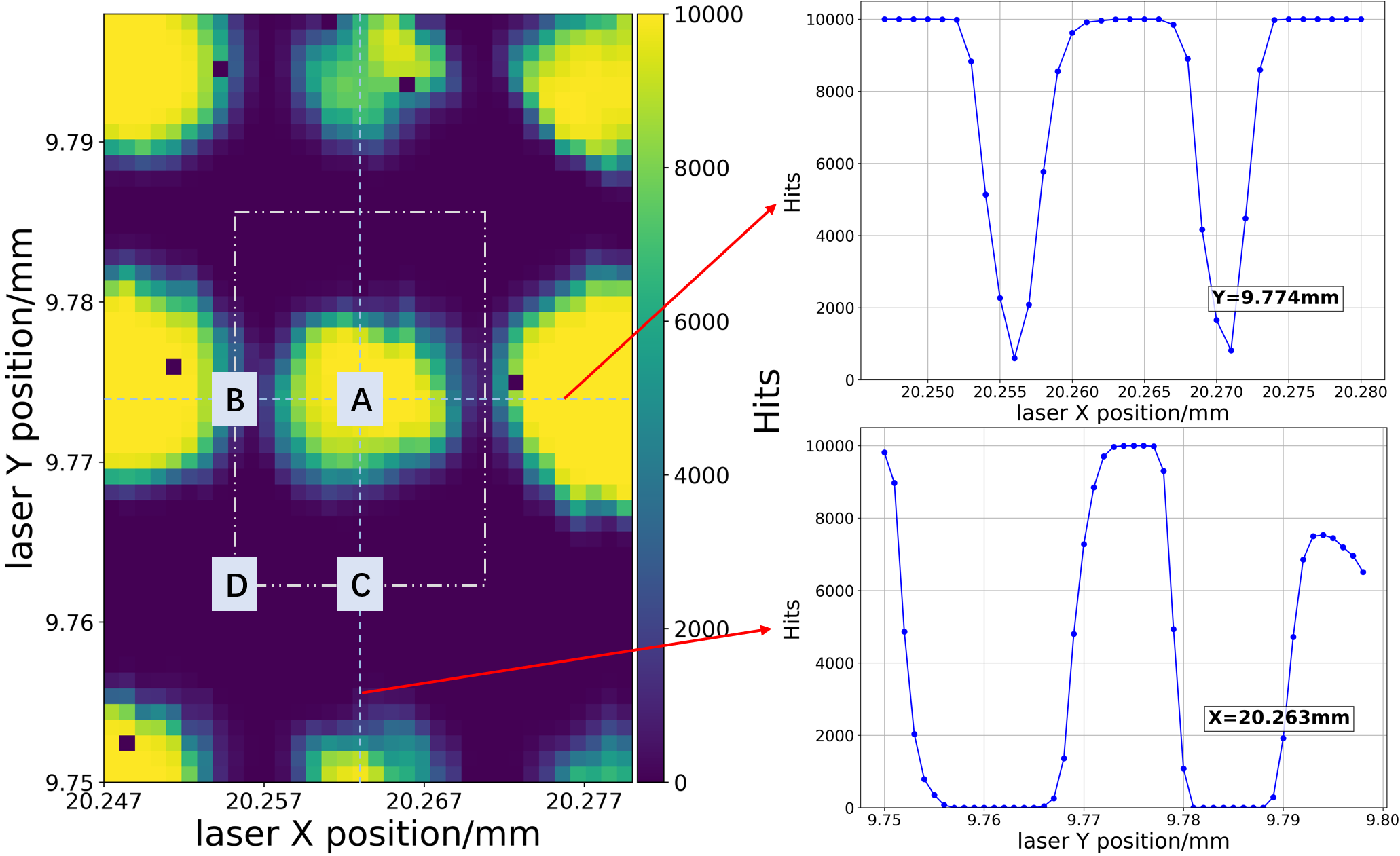}
    \caption{XY-scan results from a laser source over a selected pixel in Sector 2 (16 $\times$ 23.11 $\rm{\upmu m^2}$). The plot on the left shows the hit map, representing the number of recorded hits at each (X, Y) position. The upper right panel displays the hit profile along the line at Y = 9.774~mm, while the lower right panel shows the hit profile along the line at X = 20.263~mm. Settings of this experiment: laser frequency = 25~kHz, sensor threshold = 215~$e^-$, $V_{\rm sub}=0{\rm V}$, Rolling Shutter frames = 10,000.}
    \label{fig:xyscan}
\end{figure}

To approach the center of a single pixel, a two-dimensional XY-scan along the sensor surface was performed, as illustrated in Fig.~\ref{fig:xyscan}. For this scan, the laser intensity was significantly attenuated. This ensures that the charge deposited by a single laser pulse is small enough that at most one pixel exceeds the detection threshold, resulting in a cluster size of 0 or 1. This method provides a precise delineation of the pixel boundaries. Since the attenuated laser intensity is very close to the pixel's threshold, this measurement is particularly sensitive to small pixel-to-pixel variations in response, which explains the observed asymmetries in Fig.~\ref{fig:xyscan}. The experimental conditions included a laser frequency of 25~kHz, a threshold of 215~$e^-$ for the sensor, a substrate bias voltage $V_{\rm sub}$ of 0~V, and a total of 10,000 Rolling Shutter frames. This scan confirmed the exact center position of a single pixel in the X- and Y-directions.

\subsubsection{In-Pixel Hit Efficiency}
Previous beam test results~\cite{jadepix3telescope} indicated a non-uniform in-pixel detection efficiency, particularly between the long and short edges of the pixel. To investigate this observation, we further probed the in-pixel performance by using an attenuation-tunable infrared laser as an accessible proxy for hit efficiency and charge collection efficiency.

Based on the XY-scan results shown in Fig.~\ref{fig:xyscan}, four distinct in-pixel positions were selected and marked: \textbf{(A)} the pixel center, \textbf{(B)} the center of the long edge, \textbf{(C)} the center of the short edge, and \textbf{(D)} the diagonal center of $2\times 2$ neighbouring pixels. These four positions, annotated in Fig.~\ref{fig:xyscan}, were representative of distinct in-pixel locations and were selected to investigate spatial variations in hit efficiency.

As the infrared laser penetrates the silicon bulk and generates electrons and holes along the way, we assessed the pixel's hit efficiency in response to it to mimic the pathways of charged particles. For this laboratory measurement, we define the hit efficiency as the probability that an incident laser pulse results in a registered hit above the threshold. The laser intensity was kept the same as in the preceding XY-scan, but the pulse frequency was lowered to 25~Hz to ensure that a maximum of one laser pulse could arrive within a single frame. A 40-second Rolling Shutter scan was performed for each reverse bias value across the four in-pixel positions. Given that 1000 laser pulse triggers are expected during this period, the hit efficiency is calculated as the ratio of the measured hit count to 1000, thereby quantifying the pixel’s ability to detect individual laser pulses. 

\begin{figure}[h]
    \centering
    \includegraphics[width=0.6\textwidth]{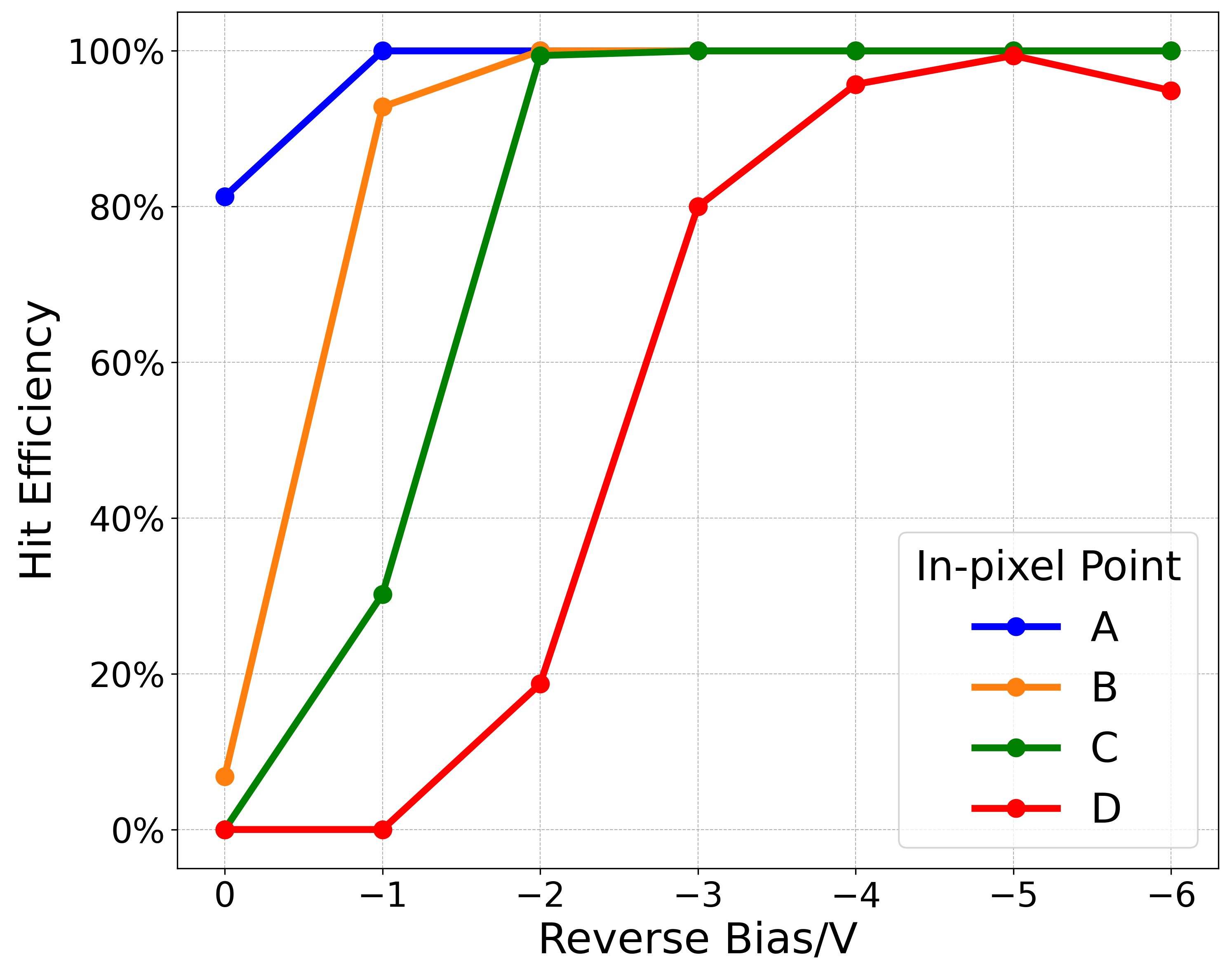}
    \caption{Hit efficiency in response to laser beam as a function of the reverse bias $V_{\rm sub}$. Measured at four in-pixel spots A, B, C, and D as marked in Fig.~\ref{fig:xyscan}. The laser intensity was the same as that used for the XY-scan, but the frequency was reduced to 25~Hz to ensure at most one pulse per frame.}
    \label{fig:eff}
\end{figure}

The results shown in Fig.~\ref{fig:eff} reveal distinct trends across the four in-pixel positions, providing information about the in-pixel variance. Position A (the pixel center) consistently exhibits the highest efficiency, while Position D (the corner) shows the lowest, reflecting diminished charge collection efficiency at the pixel boundary. Position B (long-edge center) and C (short-edge center) display intermediate efficiencies. Notably, the efficiency at Position B is higher than at Position C. This aligns with findings from the JadePix-3 beam test measurement~\cite{jadepix3telescope}, which also revealed a higher hit efficiency at the long-edge center compared to that at the short-edge center. Furthermore, the hit efficiency increases significantly with reverse bias across all positions. Applying a bias of just –2~V is sufficient to raise the efficiency at the center and edge positions (A, B, and C) to nearly 100\%, underscoring the performance improvement enabled by the expanding depletion volume. It is also noteworthy that an efficiency drop is observed at Position D when the bias increases from -5 V to -6 V. This may be caused by a smaller lateral depletion region under such a high bias condition, highlighting the importance of carefully selecting the operational bias voltage.

\subsection{Optimal Reverse Bias Setting}
A comprehensive survey of the JadePix-3’s performance over substrate biases from 0 to –6~V indicates that, under the tuning method presented in this paper, a reverse bias of approximately –5~V offers the best overall trade-off.
This optimal working-point not only improves hit efficiency and charge collection ability, but also maintains a low fake-hit rate and input capacitance.

\section{Conclusions and Outlook}\label{sec:conclusion}
To further characterize the performance of JadePix-3, a prototype MAPS developed for the CEPC vertex detector, detailed laboratory-based measurements were performed, in particular with substrate reverse bias voltages applied. A systematic test workflow was implemented for the characterization of JadePix-3.
Experimental results show that the JadePix-3 sensor can work stably when a substrate reverse bias is applied. Moreover, the reverse bias has a considerable influence on the performance of the sensor, typically being beneficial. Our results reveal that applying reverse bias substantially enhances sensor performance, reducing input capacitance, improving charge collection efficiency, and lowering fake-hit rates, though care must be taken as performance degradation, such as a drop in efficiency, was observed at the highest tested bias of -6 V. An optimal reverse bias setting is also determined within this test procedure. This test methodology confirms JadePix-3’s potential for further performance improvement.

These results acquired in laboratory conditions will also provide critical insights to inform subsequent beam test configurations, parameter optimization, as well as the radiation hardness test. Moreover, this work will serve as the architectural and procedural foundation for the tuning and optimization of the next sensor in the JadePix series, thereby driving forward MAPS development to meet CEPC’s requirements, including the ultimate 3~$\rm{\upmu m}$ spatial resolution target. Achieving this will require further dedicated R\&D, commencing with extensive beam tests to quantify the impact of the reverse-bias settings explored in this study on the spatial resolution. A critical part of this future work will be to pinpoint the root causes of degradation in position resolution under different reverse bias voltages—whether stemming from the chip design, experimental setup, or analysis algorithms—in order to guide the next sensor iteration.

\section*{Acknowledgement}
This work was supported in part by the National Key Research and Development Program of China under Grant 2022YFA1602203 and Grant 2023YFA1606300, the Fundamental Research Funds for the Central Universities of China under Grant WK2360000014, and the STCF Key Technology Research and Development Project. We thank the Hefei Comprehensive National Science Center for its strong support on the promotion and development of the STCF project.



\end{document}